\documentclass[showpacs,twocolumn,prl,citeautoscript,superscriptaddress]{revtex4-1} 
\usepackage{siunitx}
\usepackage{multirow}
\usepackage{braket}
\usepackage{booktabs}
\usepackage{graphicx}

\usepackage[table]{xcolor}

\usepackage[para,online,flushleft]{threeparttable}
\usepackage{hyperref}
\AtBeginDocument{
\heavyrulewidth=.08em
\lightrulewidth=.05em
\cmidrulewidth=.03em
\belowrulesep=.65ex
\belowbottomsep=0pt 
\aboverulesep=.4ex
\abovetopsep=0pt
\cmidrulesep=\doublerulesep
\cmidrulekern=.5em
\defaultaddspace=.5em
}

\usepackage{bm}
\usepackage{mhchem}

\begin{document}

\title{\huge Cool molecular highly charged ions for precision tests of fundamental physics}
\date{\today}
\author{Carsten Z\"ulch}
\affiliation{Fachbereich Chemie, Philipps-Universit\"{a}t Marburg, Hans-Meerwein-Stra\ss{}e 4, 35032 Marburg, Germany}
\author{Konstantin Gaul}
\email[]{konstantin.gaul@staff.uni-marburg.de}
\email[]{kongaul@web.de}
\affiliation{Fachbereich Chemie, Philipps-Universit\"{a}t Marburg, Hans-Meerwein-Stra\ss{}e 4, 35032 Marburg, Germany}
\author{Steffen M. Giesen}
\affiliation{Fachbereich Chemie, Philipps-Universit\"{a}t Marburg, Hans-Meerwein-Stra\ss{}e 4, 35032 Marburg, Germany}
\author{Ronald F. Garcia Ruiz}
\affiliation{Massachusetts Institute of Technology,~Cambridge,~MA~02139,~USA}
\affiliation{CERN, CH-1211 Geneva 23, Switzerland}
\author{Robert Berger}
\email[]{robert.berger@uni-marburg.de}
\affiliation{Fachbereich Chemie, Philipps-Universit\"{a}t Marburg, Hans-Meerwein-Stra\ss{}e 4, 35032 Marburg, Germany}
\affiliation{CERN, CH-1211 Geneva 23, Switzerland}
\maketitle
\textbf{
Molecules and atomic highly charged ions provide powerful low-energy
probes of the fundamental laws of physics: Polar molecules possess
internal fields suitable to enhance fundamental symmetry violation by
several orders of magnitudes, whereas atoms in high charge states can
feature large relativistic effects and compressed level structures,
ideally posed for high sensitivity to variations of fundamental
constants. Polar, highly charged molecules could benefit from both:
large internal fields and large relativistic effects. However, a high
charge dramatically weakens chemical bonding and drives systems to the
edge of Coulomb explosion.  Herein, we propose multiply-charged polar
molecules, that contain actinides, as promising candidates for
precision tests of physics beyond the standard model. Explicitly, we
predict \ce{PaF^{3+}} to be thermodynamically stable, coolable and
well-suited for precision spectroscopy. The proposed class of
compounds, especially with short-lived actinide isotopes from the
territory of pear-shaped nuclei, has potential to advance our
understanding of molecules under extreme conditions, to provide a
window into unknown properties of atomic nuclei, and to boost
developments in molecular precision spectroscopy in various areas,
such as optical clocks and searches for new physics.
}

As of today, important aspects of our universe are hardly understood,
such as the nature of dark matter \cite{spergel:2015} and the origin
of the imbalance between matter and anti-matter \cite{canetti:2012}.
Thus, theories that go beyond the current standard model of particle
physics are invoked, usually referred to as new physics. Such new
physics introduces additional sources of symmetry violations, like the
simultaneous violations of the symmetries with respect to spatial
inversion (known as parity $P$) and of the relative direction of time
(known as time-reversal $T$) \cite{gross:1996}. Polar, heavy-elemental
molecules like ThO allow currently some of the most precise low-energy
tests of $P,T$-violation \cite{andreev:2018} as they are easy to
polarize and possess large internal fields that enhance effects of new
physics by several orders of magnitude compared to atoms
\cite{demille:2015}. Complementary opportunities to probe new physics
are offered by atomic highly charged ions (HCIs) \cite{kozlov:2018}.
In these systems the electronic spectra are often compressed due to
the deshielded nuclear charge which results in energetically
close-lying levels as well as large relativistic effects.  This
special electronic structure can for instance provide favourable
enhancement of hypothetical spatio-temporal variations of the fine
structure constant \cite{uzan:2003} by several orders of magnitude
\cite{berengut:2010}. In contrast to neutral systems, HCIs can be
trapped comparatively easily in deep potential wells and cooled by
different mechanisms such as sympathetic cooling
\cite{drewsen:1998,bowe:1999} with well understood atomic ions such as
Be$^+$, paving the way to high-precision experiments and additionally
to direct laser-cooling \cite{kozlov:2018}. 

Small polar molecular HCIs (PMHCI) can combine advantages of polar
molecules and atomic HCIs. In addition to the unification of large
relativistic effects of HCIs with large internal fields of polar
molecules, as HCIs the PMHCIs could induce a transition from Madelung
to Coulomb ordering of electronic levels. Coulomb ordered levels could
pave the way for first direct laser-cooling of a molecular ion, as
until now laser-cooling of molecular ions was limited due to
unfavorable arrangement of electronic states compared to neutral
molecules \cite{nguyen:2011,ivanov:2020}. However, several hurdles
have to be overcome: i) Few \emph{stable} and \emph{meta-stable} small
molecules with charge number larger than two are known
\cite{schroder:1999,franzreb:2004}. Most long-lived \emph{meta-stable}
triply and quadruply charged polar diatomic molecules, which were
proposed theoretically or observed experimentally, are fluorides,
oxides or nobel gas compounds of metals
\cite{schroder:1999,franzreb:2004}. But only one \emph{stable} triply
charged polar diatomic molecule, \ce{UF^{3+}}, is experimentally
confirmed as of yet \cite{schroder:1999a}. The difficulty lies in the
requirement of very stable bonds that are able to counter the large
electrostatic repulsion of two or more positive charges that usually
lead to spontaneous Coulomb explosion. 
ii) Heavy-elemental molecules are preferred for the search for new
physics, which severely limits the choice of possible systems, as
$P,T$-violating effects are relativistic in nature and scale steeply
with increasing nuclear charge $Z$ \cite{khriplovich:1997}, Moreover,
actinide nuclei such as Pa ($Z=91$) are predicted to enhance the sensitivity to
$P,T$-violating nuclear properties by up to 5 orders of magntiude when
compared to molecules with stable nuclei \cite{chupp:2019}. iii) For precision
spectroscopy, it is is essential to cool and perfectly control the
molecule. iv) And finally a simple electronic structure is desired to
minimize systematic effects.

In this article we demonstrate that a variety of small PMHCIs suitable
for precision tests of fundamental physics exist and propose
PaF$^{3+}$ as a promising candidate. We show its stability with
respect to Coulomb explosion, characterize its electronic levels,
propose how to cool it for precision spectroscopy, and compute its
sensitivity to new physics. Moreover, we point out further possibly
stable highly charged molecules and anticipate the impact that our
results will have on molecular precision experiments in future.

The stability of a triply charged molecule \ce{AX^{3+}} can be deduced
from a simple rule of thumb \cite{bates:1955,schroder:1999}: The third
ionization energy $E_{\mathrm{i}}\left(\ce{A^2+}\right)$ of atom A
should be lower or at least nearly equal to the first ionization
energy $E_{\mathrm{i}}(\ce{X})$ of atom X. Indeed, the third
ionization energies of the first five actinides (Ac, Th, Pa, U and Np)
are low compared to those of other elements
($<\SI{20}{\electronvolt}$) \cite{migdalek:2007,wyart:1981,cao:2003}.
The first ionization energies of neon (\SI{21.564}{\electronvolt}),
fluorine (\SI{17.4}{\electronvolt}) \cite{edlen:1969} oxygen
(\SI{13.6}{\electronvolt}) and nitrogen (\SI{14.5}{\electronvolt}) are
relatively large in
comparison \cite{edlen:1969,eriksson:1963,eriksson:1971}. Whereas
trications with rare gas atoms such as \ce{CNe^3+} have been
considered theoretically early on by Koch and Frenking
\cite{koch:1987}, corresponding bond dissociation energies are
comparatively low.  Instead, we expect actinide fluorides to be most
stable followed by nitrides and oxides. \ce{UF^3+} was already shown
to be stable \cite{schroder:1999a}, but it has two unpaired
f-electrons, which may complicate the extraction of fundamental
parameters from precision spectroscopy experiments. As uranium has a
higher ionization energy than Ac, Th and Pa we can expect that the
molecules \ce{AcF^3+}, \ce{ThF^3+} and \ce{PaF^3+} are stable as well.
In the following we focus on \ce{PaF^3+} as this molecule is
isoelectronic to RaF, which has a comparatively simple electronic
structure and is known to be well suited for the study of fundamental
physics \cite{isaev:2010,isaev:2013,garciaruiz:2020,udrescu:2021}.
Moreover, the isotope $^{229}$Pa attained much attraction as it is
supposed to possess an extraordinary large static octupole
deformation \cite{ahmad:1982,ahmad:2015,singh:2019}, which would render
$^{229}$Pa highly powerful for the search for $P,T$-violations in the
quark-sector \cite{auerbach:1996,flambaum:2019}. Until now,
experimental knowledge of Pa isotopes is scarce, and molecules
containing Pa isotopes are promising systems to access electroweak
properties of Pa nuclei. To our knowledge no molecule containing
$^{229}$Pa that is suitable for precision spectroscopy was proposed so
far. Here, we show that \ce{PaF^3+} offers a versatile laboratory for
precision studies of fundamental physics.

\begin{figure*}
\includegraphics[width=\textwidth]{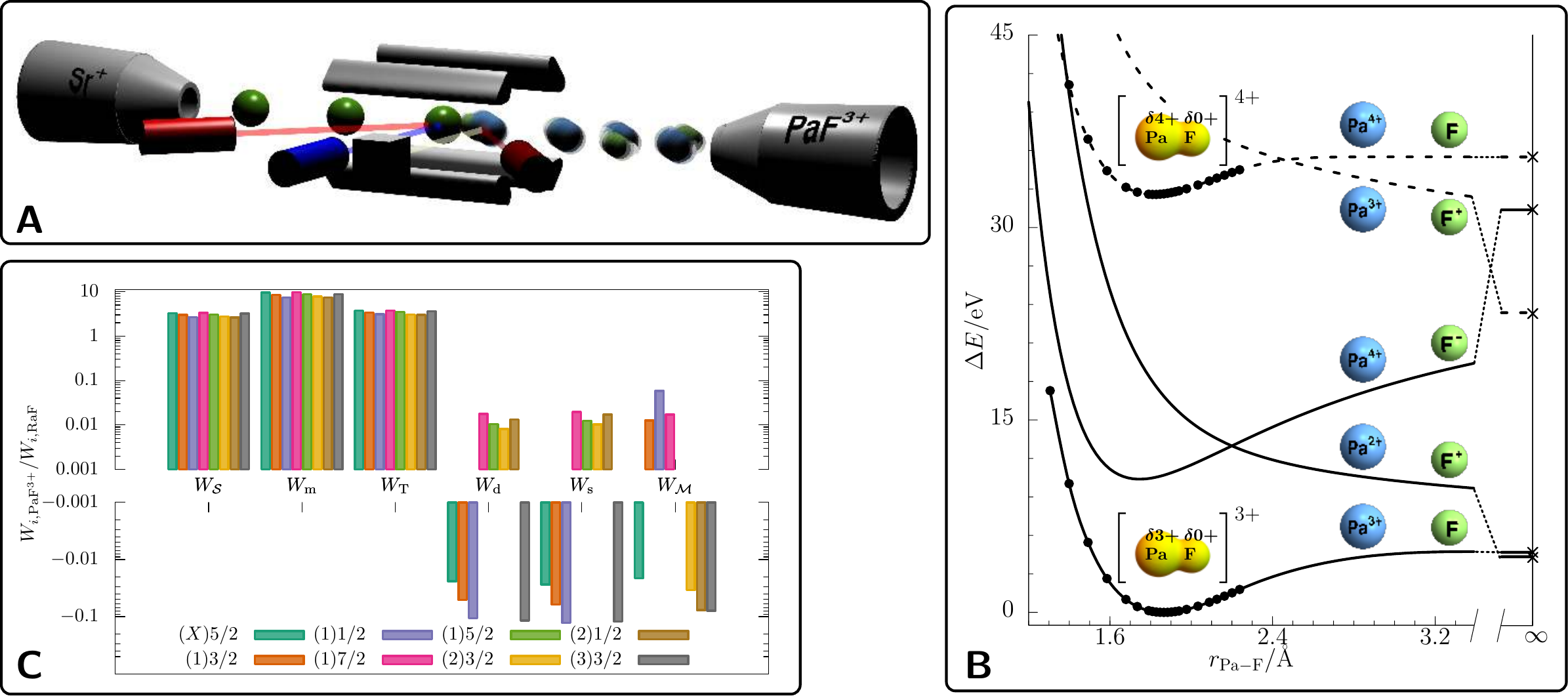}
\caption{Schematics of a search for $P,T$-violation in the
quark-sector with the stable highly charged molecule \ce{PaF^3+}.
(\textbf{A}) Sketch of an experimental set-up for precision spectroscopy with
\ce{PaF^3+}. A decelerated beam of \ce{PaF^3+} is trapped in a Paul
trap, sympathetically cooled by \ce{Sr+} ions and probed by the dark
red laser. Further lasers for a potential direct cooling of the
molecular ion \ce{PaF^3+} are omitted for clarity. (\textbf{B}) Sketch of
dissociation channels of \ce{PaF^3+} (solid lines) and \ce{PaF^4+}
(dashed lines) as diabatic potentials. Ground state electronic
potentials of \ce{PaF^3+} and \ce{PaF^4+} are shown as a spline
interpolation of points computed at the level of ZORA-cGKS-PBE0,
indicated as circles (molecular calculations) and crosses (separate
atomic calculations). Potential curves for the repulsive channels
\ce{Pa^2+} $+$ \ce{F+} and \ce{Pa^3+} $+$ \ce{F+} as well as the
attractive channel \ce{Pa^4+} $+$ \ce{F-} are represented with the
model described in the methods section. Total electronic densities and
Mulliken partial charges computed at the level of ZORA-cGKS-PBE0 are
shown for the electronic ground states.  (\textbf{C}) The parameters
of the $P,T$-odd spin-rotational Hamiltonian of \ce{PaF^3+} at the
level of ZORA-cGHF relative to those of RaF computed in
Ref.~\cite{gaul:2020} scaled by $6/1.16$ (for details see the
methods section).} 
\label{fig: setup_diss}
\end{figure*}

We study the stability of \ce{PaF^3+} with respect to the dissociation
into \ce{Pa^2+} $+$ \ce{F+}, \ce{Pa^3+} $+$ F and \ce{Pa^4+} $+$
\ce{F-} with state-of-the-art coupled cluster calculations and
quasi-relativistic density functional theory. The charge
separation dissociation channel Pa$^{2+}$ $+$ F$^+$ is at
\SI{4.1}{eV}, whereas the dissociation channels \ce{Pa^3+} $+$ F and
\ce{Pa^4+} $+$ \ce{F-} lie above this at \SI{4.9}{\electronvolt} and
\SI{32.1}{\electronvolt}, respectively. More details can be found in
the methods section.  An overview of the dissociation channels of
\ce{PaF^3+} and \ce{PaF^4+} is shown in Fig.~\ref{fig: setup_diss}B
and in Tables \ref{tab:diss_channel} to \ref{tab:paf4+}. We can conclude that
\ce{PaF^3+} is very stable.  Crude estimates of the repulsive
potential for \ce{Pa^3+} $+$ \ce{F+} suggest that even \ce{PaF^4+}
could be meta-stable, i.e.\ the charge separation channel lies below
the ground state potential but the potentials cross far from
equilibrium, with a dissociation barrier $>\SI{1}{\electronvolt}$ (see
methods section for details).

\begin{figure*}
\includegraphics[width=\textwidth]{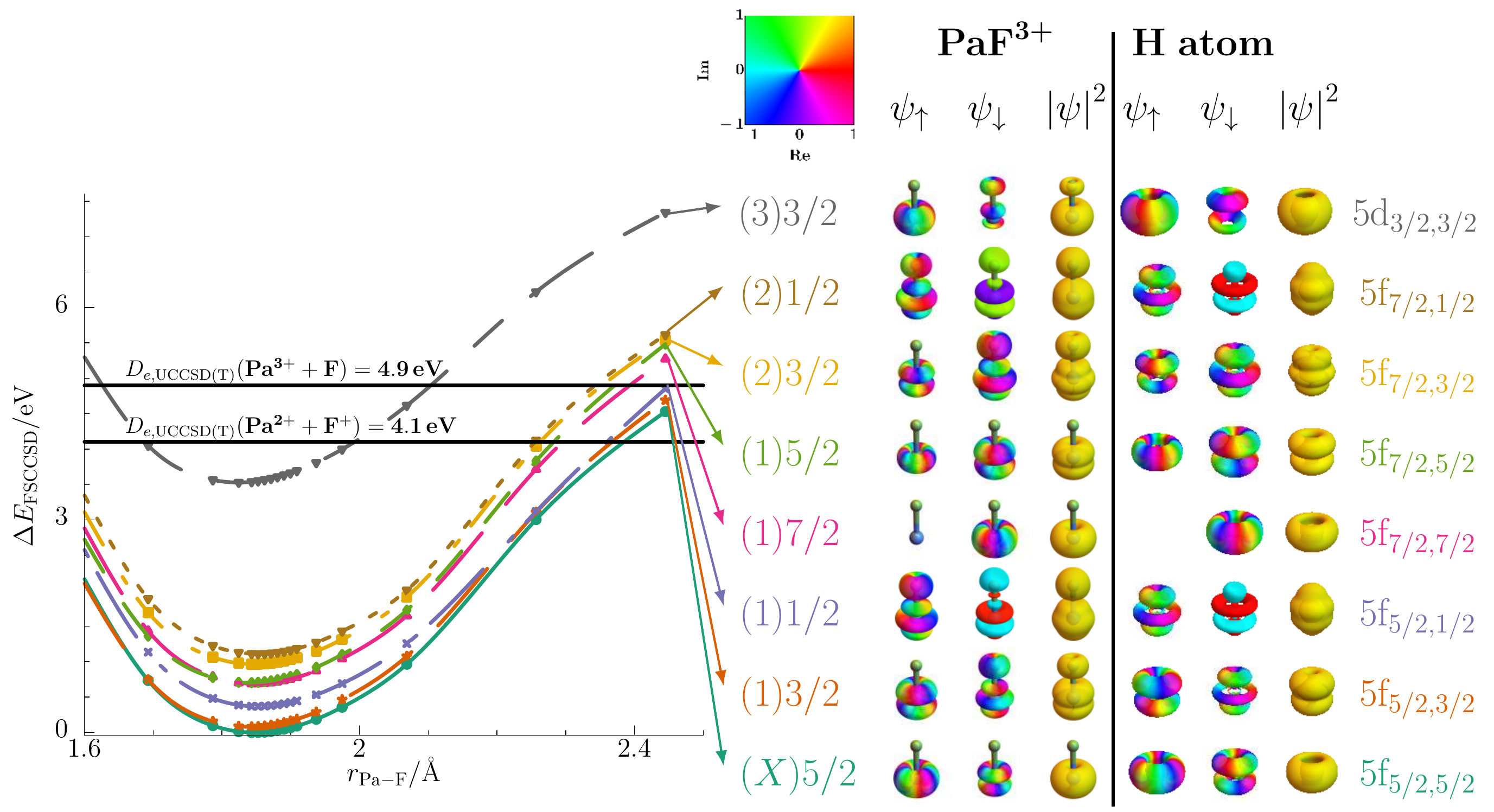}
\caption{Potential energy curves for the eight energetically lowest
electronic states relative to the ground state of \ce{PaF^3+} computed
at the level of DC-FSCCSD/ANO-RCC. Lines between the points are shown
to guide the eye. The two lowest dissociation channels \ce{Pa^2+} $+$
\ce{F+} and \ce{Pa^3+} $+$ \ce{F}, as computed on a highler level of
theory (RECP-UCCSD(T)-SOC), are indicated by black horizontal lines.
The various electronic states are additionally characterized by the
complex two-component ZORA-cGHF spinor
$\psi=(\psi_\uparrow,\psi_\downarrow)$ of the unpaired electron and
compared to the upper component of the analytic solution of the Dirac
equation for the hydrogen atom. Atomic hydrogen spinors are labelled
as $\ell_{j,m_j}$, where $\ell$ is the symbol for the electronic
orbital angular momentum quantum number, $j$ is the total electronic
angular momentum quantum number and $m_j$ is the magnetic total
electronic angular momentum quantum number.}
\label{fig: potentials_orbitals}
\end{figure*}

We computed vertical excitation energies, equilibrium bond lengths and
harmonic vibrational wavenumbers of \ce{PaF^3+} on the level of
Dirac--Coulomb Fock-Space Coupled Cluster with Singles and Doubles
amplitudes (DC-FSCCSD) and within a Zeroth Order Regular Approximation
complex Generalized Hartree-Fock (ZORA-cGHF) self consistent field
maximum overlap approach. With the ZORA-cGHF approach we determine
transition electric dipole moments, projection of the orbital angular
momentum on the molecular axis and hyperfine coupling constants, here
reported for $^{231}$PaF$^{3+}$, as nuclear magnetic dipole moments
for other Pa isotopes are lacking.  ZORA-cGHF calculations can be
assumed to be accurate within about $10\,\%$ in comparison to FSCCSD
calculations. And effects of larger basis sets are on the order of
$5\,\%$ (for details see the methods section and Table
\ref{tab:spectroscopy}).

All electronic states explicitly studied herein are below
the lowest dissociation channel at \SI{4.1}{eV} (see Fig.~\ref{fig:
potentials_orbitals} on the left and Table \ref{tab:spectroscopy}; potential
curves at other levels of theory are provided in Supplementary Figures
S1 -- S5). A short bond length of about
$\SI{1.85}{\angstrom}$ and a large harmonic vibrational wavenumber
$\sim\SI{850}{\per\centi\metre}$ indicate a strong Pa--F bond. The
level of mixing due to spin-orbit coupling is estimated from the
projection of the orbital angular momentum on the molecular axis
$\Lambda$. We find that the $(X)5/2$ ground state is of $90\,\%~\Phi$
and $10\,\%~ \Delta$ character and the first two excited states are of
$80\,\% ~\Delta$ and $20\,\%~\Pi$ character [$(1)3/2$ state] and
$60\,\% \Pi$ and $40\,\%~\Sigma$ character [$(1)1/2$ state],
respectively. As can be seen in Fig.~\ref{fig: potentials_orbitals} on
the left all electronic states of \ce{PaF^3+} appear to have almost parallel
potential curves, with bond lengths and harmonic vibrational
wavenumbers varying less than $\SI{5}{\percent}$. This indicates that
electronic excitations are determined by transitions between
non-bonding spinors or spinors with nearly the same bonding or
anti-bonding character.  To characterize the involved atomic orbitals
we compare the complex singly occupied ZORA-cGHF spinor to the large
component of the analytic solution of the Dirac equation for the
hydrogen atom for principal quantum number $n=5$ (see Fig. \ref{fig:
potentials_orbitals} on the right). This confirms that the excited states are
dominated by single-electron transitions between non-bonding or weakly
anti-bonding orbitals. The seven lowest states are dominated by a
singly occupied f-orbital located at Pa, of which the three lowest
electronic states are characterized by a $5\mathrm{f}_{5/2}$ orbital,
and the next four states are characterized by a $5\mathrm{f}_{7/2}$
orbital. 

The lowest six transition energies are narrowly spaced and squared
transition electric dipole moments $\left|\vec{\mu}\right|^2$ indicate
low transition rates.  This is in accordance with electronic
transitions involving primarilly f-type atomic orbitals.  Einstein
coefficients for spontaneous emission roughly estimated from
$\left|\vec{\mu}\right|^2$ and $T_\mathrm{e}$ at the level of
ZORA-cGHF considering only electronic degrees of freedom are provided
in Table \ref{tab:S7}. From these we infer that the radiative
lifetimes of the three lowest electronically excited states [$(1)3/2$,
$(1)1/2$, $(1)7/2$] could be on the order of \si{\milli\second}, which
is on the same order as the lifetime of the $H^3\Delta_1$ state in ThO
that was used to provide the so far tightest upper bounds on molecular
$P,T$-violation \cite{andreev:2018}. \ce{PaF^3+} can be trapped and,
due to a mass to charge ratio of about $83$~u/e, efficiently cooled
sympathetically with \ce{Sr+} ions as sketched in
Fig.~\ref{fig: setup_diss}A.  Moreover, other cooling schemes, for
instance with buffer gases, can be considered. In addition we find
almost diagonal Franck--Condon matrices for all electronic transitions
(see Table \ref{tab:S7}) because of the non-bonding or weakly
anti-bonding character of the highest occupied spinor and consequently
almost parallel potential curves. When combined with an efficient
pre-cooling scheme, direct laser-cooling of \ce{PaF^3+} seems
feasible. For instance the $(1)7/2\leftarrow (X)5/2$ transition, which
is at about \SI{1800}{\nano\metre} has an estimated cumulated
Franck--Condon factor of 0.999997 when taking the 0-0, 1-0 vibrational
transitions into account on this level of theory and an estimated
lifetime of $<\SI{20}{\milli\second}$.  Another possibility for direct
laser-cooling could be a population of the meta-stable $(1)1/2$-state
and cycling in the $(2)1/2\leftarrow (1)1/2$ transition at about
\SI{1800}{\nano\metre}, which has a similar estimated Franck--Condon
factor (0.999992) but a probably much shorter lifetime
(\SI{<70}{\micro\second}). These properties clearly point to
favourable prospects to obtain cold samples of \ce{PaF^3+} for
precision experiments.

\begin{figure*}
\includegraphics[width=\textwidth]{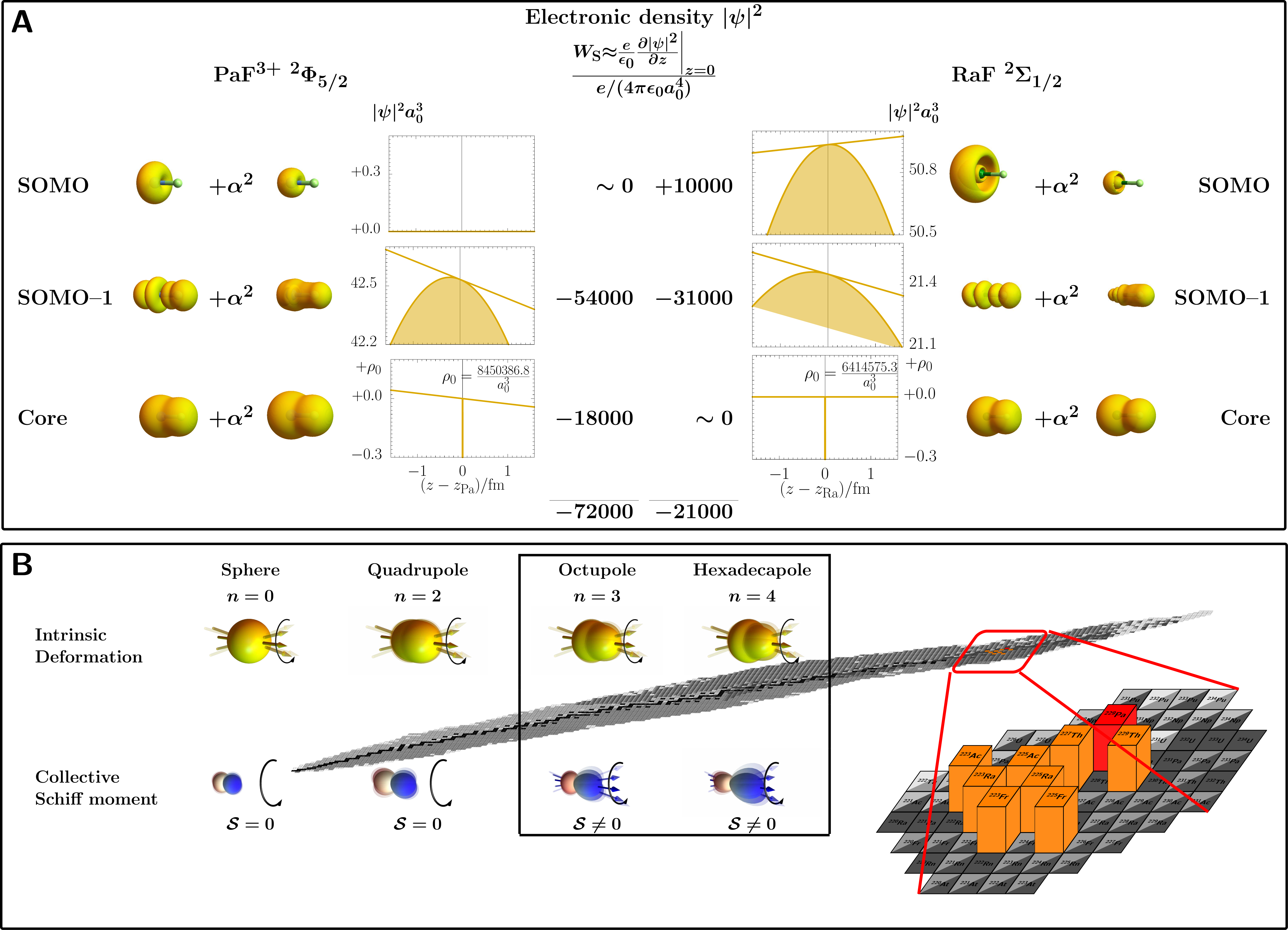}
\caption{Visualization of enhancement of $P,T$-violation in the
quark-sector within the PMHCI \ce{PaF^3+} compared to isoelectronic
RaF, illustrated by example of the nuclear Schiff moment. This moment
would induce an energy shift propotional to
$\mathcal{S}W_\mathcal{S}\mathcal{I}$ (see text). (\textbf{A})
Enhancement as per the electronic structure parameter $W_\mathcal{S}$
is shown. Non-vanishing $W_\mathcal{S}$ stems from a shift of the
maximum electron density away from the nuclear center by polarization,
which in good approximation is characterized by the slope of the
four-component (4c) electron density (indicated by a line in the plots
in the middle) at the position of the Pa or Ra nucleus.  Contributions
from the SOMO, the highest doubly occupied orbital (SOMO--1) and
uncompensated contributions from core electrons are shown. Approximate
4c electron densities computed at the ZORA-cGHF level are plotted with
contour value of $0.035^2~a_0^{-3}$, with lower component density
contributions (right isodensities) being magnified by $\alpha^{-2}$ to
enhance visibility as compared to the upper component (left
isodensities).  In each isodensity plot, the heavy nucleus (Pa, Ra) is
located on the left, the fluorine nucleus on the right as indicated by
the ball-and-stick structures. (\textbf{B}) Visualization of the
strong enhancement of a collective Schiff moment $\mathcal{S}$ in
octupole deformed nuclei. Large octupole deformations accumulate in
the highlighted region of the nuclide chart. Nuclear charge densities
with deformation of order $n$ and corresponding angular densities of
the Schiff operator are modeled in spherical plots (for details on how
this is realized see the methods section). The resulting collective
Schiff moments $\mathcal{S}$ are indicated by a vector of length
$\mathcal{S}$ along the deformation axis in the intrinsic frame. The
precession of the intrinsic moment around the $z$-axis is also
indicated.} 
\label{fig: schiff_moment}
\end{figure*}

In order to estimate the enhancement of new physics effects in
\ce{PaF^3+} we compute the electronic structure parameters $W_i$ of the
$P,T$-violating spin-rotational
Hamiltonian \cite{hinds:1980,kozlov:1995,gaul:2020} that reads in good
approximation (see methods section for details) 
\begin{equation}
\begin{aligned}[t]
H_{\mathrm{sr}} &=\Omega
\left(W_{\mathrm{d}}d_\mathrm{e}+W_{\mathrm{s}}
k_\mathrm{s}
\right)
+\Theta
W_{\mathcal{M}}\tilde{\mathcal{M}} \\
&+\mathcal{I}
\left(
W_{\mathrm{T}} k_\mathrm{T}
+W_{\mathcal{S}} \mathcal{S}
+(W_{\mathrm{m}}+W_{\mathcal{S}}R_\text{vol}) d_\mathrm{p}
\right)\,,
\end{aligned}
\label{eq: ptodd_spinrot}
\end{equation}
where $\Omega$ is the projection of the total electronic angular
momentum on the molecular axis, $\mathcal{I}$ is the projection of
total spin of Pa on the molecular axis and $\Theta$ accounts for the
electron and nuclear spin interaction along the molecular
axis \cite{kozlov:1995,gaul:2020}. We account here for $P,T$-violation
in a molecule via an electric dipole moment of the electron
$d_\mathrm{e}$, an electric dipole moment of the proton
$d_\mathrm{p}$, the collective Schiff moment $\mathcal{S}$, the
nuclear magnetic quadrupole moment $\mathcal{M}$, the $P,T$-odd
scalar-pseudoscalar nucleon-electron current $k_\mathrm{s}$, and
tensor-pseudotensor nucleon-electron current $k_\mathrm{T}$
interactions. $R_\text{vol}$ is a nuclear structure factor.  \emph{Ab
initio} results for the various $W_i$ parameters at the level
ZORA-cGHF for the eight lowest electronic states of \ce{PaF^3+} are
provided in Table \ref{tab:properties} and are compared to the
isoelectronic RaF molecule in Fig.~\ref{fig: setup_diss}C. The large enhancement of $P,T$-odd effects that stem from
$P,T$-violation in the quark-sector.  In the $X(5/2)$ state the
collective Schiff moment is enhanced by
$W_\mathcal{S}\sim\SI{-72000}{\elementarycharge/(4\pi\epsilon_0)\per\bohr^4}$,
which is more than three times larger in absolute value than in
isoelectronic RaF
($\sim\SI{-21000}{\elementarycharge/(4\pi\epsilon_0)\per\bohr^4}$
computed with the same method \cite{gaul:2020}). We analyse this large
absolute value of $W_{\mathcal{S}}$ in comparison to RaF in
Fig.~\ref{fig: schiff_moment}.  Whereas in RaF the singly occupied
molecular orbital (SOMO) has a pronounced s-character, with its
contribution to $W_{\mathcal{S}}$ partially cancelling the
contribution from the highest doubly occupied orbital (SOMO-1), there
is no contribution from the f-type SOMO in \ce{PaF^3+}. Furthermore,
we see an additive uncompensated contribution from the core orbitals
in \ce{PaF^3+} and a much larger contribution from the (SOMO-1), which
can be attributed to the pronounced relativistic effects in
\ce{PaF^3+}.  Similar effects are observed for the enhancement factors
$W_{\mathrm{T}}$ and $W_{\mathrm{m}}$. Thereby, the magnetic
interaction with a valence proton $W_{\mathrm{m}}$ is up to ten times
larger than in RaF. We can thus conclude that \ce{PaF^3+} has a
pronounced sensitivity to $P,T$-violation in the quark-sector. 

Atomic f-type orbitals, which determine essentially the SOMO of
\ce{PaF^3+}, have vanishing probability density within the nucleus, so
that $P,T$-odd effects dependent on the electron spin ($W_\mathrm{d}$,
$W_\mathrm{s}$) are suppressed. However, in the possibly meta-stable
$(1)1/2$-state, these effects are still on the same order as in BaF,
for which an experiment to search for $d_\mathrm{e}$ is
planned \cite{aggarwal:2018}. The relative suppression of electron-spin
dependent effects can be an advantage for the disentanglement of the
fundamental sources of $P,T$-violation, for which the ratio of
different enhancement factors $W_i$ plays an important
role \cite{chupp:2019,gaul:2019}. The ratio of
$W_\mathrm{d}/W_\mathrm{s}$ is $12\,\%$ lower than predicted by the
model presented in Ref.~\cite{gaul:2019} for
$Z=91$ and, thus, would correspond to $Z=97$ in the model.  This can
be explained by the high charge of the molecule that leads to a higher
effective nuclear charge. The ratios between other parameters are
considerably different compared to RaF and some even have opposite
sign. All this renders experiments with
\ce{PaF^3+} complementary in the search for $P,T$-violation to other
experiments with polar open-shell molecules. 

Moreover, \ce{PaF^3+} can also be advantageous for other tests of
fundamental physics. The electronic potentials of
\ce{PaF^3+} are largely overlapping, leading to possibly very close
lying nearly degenerate vibrational states. In combination with the
large relativistic effects this could result in a high
sensitivity to a variation of fundamental constants \cite{chin:2009}.

In summary, highly charged polar actinide molecules can be stable and
have favourable properties for precision tests of fundamental physics.
Such PMHCIs can be cooled sympathetically and can show a pronounced
enhancement of new physics effects in the quark-sector. In particular,
we demonstrated that the molecule \ce{PaF^3+} is favorable for
precision tests of fundamental physics.  \ce{PaF^3+} has a very simple
electronic structure with one valence f-electron in a non-bonding or
weakly anti-bonding orbital located at Pa. This leads to parallel
potential curves for essentially all energetically low-lying
electronic states, which may allow to establish an efficient direct
laser-cooling scheme after sympathetic pre-cooling with \ce{Sr+} ions
or pre-cooling in a buffer-gas cell. Furthermore, due to the high
charge many close lying states can be found that may be advantageous
in the search for a variation of fundamental constants.

Beyond \ce{PaF^3+}, we suggest to study also the PMHCIs \ce{UO^3+} and
\ce{NpN^3+} as possibly stable candidates for molecular precision spectroscopy.
From the discussion above we infer, moreover, that \ce{ThF^3+} can be a very
stable closed-shell highly charged molecule.  Furthermore, doubly charged
molecular ions, such as \ce{ThF^2+} and \ce{PaO^{2+}} that are isoelectronic to
RaF, might also be considered for high-precision spectroscopy.  Some selected
molecular properties computed for \ce{PaO^{2+}} are reported in Table~S6-S8 in
the Supplementary Material.  Previously, \ce{PaO^{2+}} was studied
theoretically \cite{kovacs:2013} and observed in the gas-phase
\cite{santos:2006}. From this it can be expected that \ce{PaO^{2+}} has similar
properties as \ce{PaF^3+}.  Sympathetic cooling or pre-cooling of these doubly
charged ions may be achievable with \ce{Ba+} ions. 

Recent progress \cite{isaev:2016,kozyryev:2016,fan:2021} allows
precision searches for new physics with polyatomic molecules and
polyatomic molecular ions which provide several experimental benefits
\cite{isaev:2017,kozyryev:2017a}. We indicate here for instance
\ce{PaNC^3+}, \ce{PaNCS^3+} as candidates for possibly stable
polyatomic PMHCIs. Following
Ref.~\cite{harvey:2003} it may be worthwhile to
study \ce{[PaNCCH3]^4+} as potentially long-lived meta-stable
symmetric-top PMHCI for a precision search for new physics.

Our study opens up a new route to molecular precision spectroscopy and
is a starting point to search for further candidates of molecular
HCIs. We anticipate that this class of systems has possible
applications as optical clocks, for quantum logic spectroscopy and for
precision test of new physics. Molecular HCIs such as \ce{PaF^3+}
advance our understanding of chemical bonding, can provide a powerful
probe of nuclear electroweak properties and will boost precision
searches for new physics beyond the standard model. These studies
provide further motivation for the emergent field of short-lived
radioactive molecules. This is timely with the development of future
radioactive beam facilities such as the Facility for Rare Isotopes
(FRIB) in the U.S., which starts operation in 2022, and is expected to
produce unprecedented amounts of Pa isotopes and other actinide
nuclei.

\emph{Acknowledgments.}---%
We thank Nicholas R. Hutzler, Andrei Zaitsevskii and Timur Isaev for
discussions and Gernot Frenking for his comments on the manuscript.
R.B.\ is indebted to Helmut Schwarz, Jana Roithova and the late Detlef
Schr{\"o}der for early discussions on highly charged molecular ions.
This work was funded by the Deutsche Forschungsgemeinschaft (DFG,
German Research Foundation) -- Projektnummer 445296313. Computer time
provided by the center for scientific computing (CSC) Frankfurt is
gratefully acknowledged.  

\emph{Author contributions.}---%
R.F.G.R.\ had the initial idea to search for a Pa containing molecule
for tests of fundamental physics. K.G and R.B.\ proposed
\ce{PaF^3+} as possible candidate molecule and coordinated the
research project. K.G.\ did preliminary calculations on the stability
of \ce{PaF^3+} and its symmetry violating properties and implemented
the MOM scheme for quasi-relativistic calculations of excited states.
S.M.G.\ provided an implementation for matrix elements of
non-orthogonal wave function and for visualization of complex
two-component orbitals. C.Z.\ did all calculations presented in the
final version of the manuscript. C.Z.\ and K.G.\ wrote the initial
draft of the manuscript. All authors contributed to the discussion and
the final version of the manuscript.

\emph{Extended Data.}---%
Explicit numerical values for dissociation energies of \ce{PaF^3+} and
\ce{PaF^4+} and ionization energies of \ce{Pa}, \ce{F} are given at
different levels of theory. Spectroscopic constant of \ce{PaF^3+} are
provided at the two-component ZORA and FSCCSD levels of theory.
Franck-Condon factors, Einstein coefficients and estimated lifetimes
for the eight lowest states of \ce{PaF^3+} are given at the level of
ZORA-cGHF. Explicit values of $P,T$-odd properties for the eight
lowest electronic states are given at the ZORA two-component level.

\emph{Supplementary information.}---%
We provide the potential energy curves of selected electronic states
of \ce{PaF^3+} at different levels of theory and the potential energy
curve of \ce{PaF^4+} at the level of FSCCSD.  Selected properties of
\ce{PaO^2+} for the lowest three electronic states are given at the
level of ZORA-cGKS-B3LYP.  We provide all data to reproduce the
figures shown in the main text and the supplement. 

\bibliographystyle{naturemag}

\clearpage
\section*{Methods}
\subsection*{Computational methods}
Relativistic four component calculations were performed with the
quantum chemistry program package \textsc{Dirac19} \cite{dirac19}
employing the Dirac--Coulomb (DC) Hamiltonian. The four-component
Dirac--Coulomb Hartree--Fock (DC-HF) method was used to describe
the electronic structure of Fluorine and Protactinium in different 
electron configurations, with the open-shell situation being 
treated within the average-of-configuration framework (ORHF).
Relativistic Fock-space coupled-cluster calculations with singles and
doubles cluster amplitudes (DC-FSCCSD) were performed starting from the
PaF\textsuperscript{4+} closed-shell electronic ground state as reference 
wavefunction and attaching an additional electron. Seven
electrons of F ($2s^2 2p^5$) and 19 electrons of Pa ($5d^{10} 6s^2
6p^6 5f^1$) were explicitly included in the electron correlation 
treatment. Virtual spinors were considered
up to an energy of \SI{50}{\hartree}. In DC-ORHF and DC-FSCCSD
calculations we employed the Dyall all-electron triple zeta (Dyall-ae3z)
basis set for the Pa and F atom \cite{dyall:2002,dyall:2006} or a
triple zeta atomic natural orbital type basis set (ANO-RCC-VTZP)
\cite{roos:2005}. To assess the quality of the basis set, additional
calculations were performed with the Dyall all-electron quadruple zeta
(Dyall-ae4z) basis set \cite{dyall:2002,dyall:2006}. DC-ORHF spinors
were optimized in a self-consistent manner until a change in the
orbital gradient below \SI{e-7}{\per\bohr^{3/2}} was reached.

Quasirelativistic two-component calculations were performed with a
modified version \cite{wullen:2010} of the quantum chemistry program
package \textsc{Turbomole} \cite{ahlrichs:1989} at the level of
complex generalized Hartree-Fock (cGHF) and complex generalized
Kohn-Sham (cGKS) density functional theory (DFT) within the zeroth
order regular approximation (ZORA) employing a model potential to
alleviate the gauge dependence of the ZORA Hamiltonian as proposed by
van W\"ullen \cite{wullen:1998}. The ZORA-model potential was employed
with additional damping \cite{liu:2002}. Calculations on the DFT level
were performed with the hybrid Becke three parameter exchange
functional and the Lee, Yang and Parr correlation functional (B3LYP)
 \cite{vosko:1980,becke:1988,lee:1988,stephens:1994}, as well as with
the hybrid version of the Perdew, Burke, Ernzerhof functional (PBE0)
 \cite{perdew:1996,adamo:1999}. On this level of theory, a basis set
consisting of 37s, 34p, 14d and 9f uncontracted Gaussian functions
with the exponential coefficients $\alpha_i$ composed as an
even-tempered series as $\alpha_i=a/b^{i-1}$; $i=1,...,N$, where $N$
is the number of functions, with $b=2$ for s and
p functions and $b=(5/2)^{1/25} \times 10^{2/5}\approx 2.6$ for d
and f functions was used for Pa. The largest exponent coefficients
$a$ of the subsets are $2\times10^9~a_0^{-2}$ (s), $5\times 10^8~a_0^{-2}$
(p), $13300.758~a_0^{-2}$ (d) and $751.8368350~a_0^{-2}$ (f). A
decontracted atomic natural orbital basis set of double-$\zeta$
quality augmented with polarization valence basis functions
(ANO-RCC-VDZP) \cite{roos:2004} was used for the F atom. These basis
sets have performed well in previous studies of $P,T$-violation in
molecules \cite{isaev:2013,gaul:2017,gaul:2019}.

In all relativistic or quasi-relativistic calculations, a normalized
spherical Gaussian nuclear density distribution $\rho_A \left( \vec{r}
\right) = \frac{\zeta_A^{3/2}}{\pi ^{3/2}} \text{e}^{-\zeta_A \left|
\vec{r} - \vec{r}_A \right| ^2}$ with $\zeta_A = \frac{3}{2 r^2
_{\text{nuc},A}}$ and the root-mean-square radius $r_{\text{nuc},A}$
was used as a finite nucleus as suggested by Visscher and Dyall
\cite{visscher:1997}. Isotopes $^{231}$Pa and $^{19}$F were used
to determine the size of the finite nucleus.
An exception to this are calculations of the energy gradient within
the ZORA-cGKS approach, which were performed assuming a pointlike
nucleus instead.

At the level of cGKS and cGHF, excited state orbitals were obtained by
SCF calculations choosing occupation numbers regarding to maximum
overlap with the determinant of the initial guess (initial guess
maximum overlap method, IMOM) \cite{gilbert:2008,barca:2018}. As
initial guess we used the cGKS or cGHF determinant, that was found
with occupation of energetically lowest spinors. If the change in the
differential density with respect to the previous cycle was below
$10^{-3}/a_0^{-3}$ the standard MOM was used, where occupation numbers
are chosen with respect to maximum overlap with the determinant of the
previous cycle \cite{gilbert:2008}.

Franck--Condon factors were obtained with the program package
\textsc{hotFCHT}
 \cite{berger:1997a,jankowiak:2007,huh:2012proc,huh:2010} by
calculating the harmonic force constants with the module NumForce
within the modified version of Turbomole mentioned above.

Transition electric dipole moments $\vec{\mu}$ were computed for the independently
obtained cGHF and cGKS determinants using L{\"o}wdin rules \cite{Lowdin:55} for
single-particle operator transition matrix elements between nonorthogonal 
single-determinantal initial wave function $\Phi_\mathrm{i}$ and final wave function 
$\Phi_\mathrm{f}$
\begin{equation}
\vec{\mu} = \Braket{\Phi_\mathrm{f} | \vec{r} | \Phi_\mathrm{i}} =
\sum_{ij} \Braket{\psi_{\mathrm{f},i} | \vec{r} | \psi_{\mathrm{i},j}} \text{adj}\left( \bm{S} \right)_{ij} 
\label{eq:loewdin}
\end{equation} 
with the molecular spinors $\psi_i$ of the initial (i) and final (f)
state, approximately described by a single Slater determinant, and the adjugate 
of the overlap matrix between the initial and final state determinant with elements
$S_{ij}=\Braket{\psi_{\mathrm{f},i}|\psi_{\mathrm{i},j}}$.

All other molecular properties were computed with the toolbox approach
presented in Ref.~\cite{gaul:2020}. We neglected magnetic and
hyperfine coupling induced interactions of $d_\mathrm{e}$ and
$k_\mathrm{s}$ as well as higher order $P,T$-odd nuclear moments and
described the nuclei in the non-relativistic limit. For a definition
of the various electronic structure factors $W_i$ except for
$W_\mathcal{S}$ we refer to Ref.\nocite{gaul:2020}\citenum{gaul:2020}.
We used here a different definition of $W_\mathcal{S}$ than in
Ref.\nocite{gaul:2020}\citenum{gaul:2020} which includes a factor of
6. Furthermore, we employed a finite nucleus model for the calculation
of the $W_\mathcal{S}$ operator \cite{flambaum:2002}, whereas in
Ref.\nocite{gaul:2020}\citenum{gaul:2020} a point-like nulcear model
was used:
\begin{equation}
\begin{aligned}
W_\mathcal{S} &= -4 \pi k_\mathrm{es} e \Braket{\Psi|\frac{\partial}{\partial
z}\rho_A(\vec{r})|\Psi} \\
&= 8 \zeta_A \pi k_\mathrm{es} e 
\Braket{\Psi|(z-z_A)\rho_A(\vec{r})|\Psi}\,,
\end{aligned}
\end{equation}
where the electrostatic constant $k_\mathrm{es}$ is
$\frac{1}{4\pi\epsilon_0}$ in SI units and $\epsilon_0$ is the
electric constant. The last equality applies only to the spherical
Gaussian nuclear model used in this work. In opposite to the operator
used in Ref.\nocite{gaul:2020}\citenum{gaul:2020} this operator can be
evaluated in numerical integration within the toolbox apporach. As
suggested in Ref.\nocite{flambaum:2020a}\citenum{flambaum:2020a}, for
direct comparison to the present values computed with a finite nuclear
model for PaF$^{3+}$, we scaled the corresponding values of
$W_\mathcal{S}$ of RaF that were obtained with the operator for a
pointlike nucleus in Ref.\nocite{gaul:2020}\citenum{gaul:2020} by a
factor of $1/1.6$.

Bond lengths were optimized to an energy change of less than
$10^{-6}~E_{\text{h}}$ as convergence criterion. The wavefunctions were
optimized to a change in energy and spin-orbit coupling contribution of
$10^{-11}~E_{\text{h}}$ or better, with exception of the calculations
with the ANO-RCC-VTZP basis set, which was optimized to a change in
energy and spin-orbit coupling contribution of $10^{-9}~E_{\text{h}}$. 

Calculations in a non-relativistic framework using scalar relativistic ECPs
(RECP), which are reported in Tables \ref{tab:diss_channel}--\ref{tab:paf4+},
were performed with the quantum chemistry program package \textsc{Molpro}
\cite{molpro2012a,molpro2019a,molpro2020} on the level of restricted open-shell
Hartree-Fock (ROHF), spin-unrestricted-Kohn-Sham (UKS) using the functionals
B3LYP and PBE0 and on the level of unrestricted coupled-cluster with iterative
singles and doubles amplitudes combined with perturbative triples amplitudes
(UCCSD(T)). For Pa a relativistic energy-consistent small-core pseudopotential
(ECP60) was used together with atomic natural orbital valence basis set
\cite{cao:2003b}. An augmented correlation-consistent polarized basis with
quadruple-$\zeta$ qualtiy (aug-cc-pVQZ) \cite{dunning:1989} was used on the
F-atom. The bond length was optimized up to a change in energy of
$10^{-6}~E_\mathrm{h}$.  Self consistent field calculations were performed
until a change in the gradient in respect to the orbital rotation lower than
$10^{-13}~E_\mathrm{h}/a_0$ was reached.

The complex two-component orbitals are visualized by calculating orbital 
amplitudes on a three-dimensional grid and plotting them
with the help of \textsc{Mathematica} version 11 \cite{mathematica11} by mapping
the phase in the complex plane via a color code on the contour surface of
the absolute value of the spinors.

\subsection*{Disscociation channels}
We study the stability of \ce{PaF^3+} with respect to the dissociation
into \ce{Pa^2+} $+$ \ce{F+}, \ce{Pa^3+} $+$ F and \ce{Pa^4+} $+$
\ce{F-} by separate energy calculations of the atomic products
and the corresponding molecular species at its equilibrium structure
with state-of-the-art unrestricted coupled cluster
calculations with single and double amplitudes and preturbative
triples [UCCSD(T)], in which we account for scalar-relativistic
effects by an relativistic effective core potential (RECP).  We find
that the charge separation dissociation channel Pa$^{2+}$ $+$ F$^+$ is
at \SI{5.1}{eV}. The dissociation channels \ce{Pa^3+} $+$ F and
\ce{Pa^4+} $+$ \ce{F-} lie above this at \SI{5.8}{\electronvolt} and
\SI{32}{\electronvolt}, respectively. In these calculations spin-orbit
coupling is, however, not accounted for. To quantify this effect,
quasi-relativistic two-component complex generalized
Kohn-Sham (ZORA-cGKS) calculations within zeroth order regular
approximation are compared to relativistic effective core potential
unrestriced Kohn-Sham calculations (RECP-UKS), using the PBE0
functional, which performed well in the computation of bond
dissociation energies of uranium halides \cite{batista:2004}. The
size of the spin-orbit coupling contribution to the three dissociation channels is
\SI{-1.0}{\electronvolt}, \SI{-0.9}{\electronvolt} and
\SI{0.1}{\electronvolt}, respectively, yielding the spin-orbit
corrected [RECP-UCCSD(T)+SOC] dissociation energies of
\SI{4.1}{\electronvolt}, \SI{4.9}{\electronvolt} and
\SI{32.1}{\electronvolt}, which are in good agreement with
dissociation energies at the level of ZORA-cGKS-PBE0
(\SI{3.8}{\electronvolt}, \SI{4.7}{\electronvolt} and
\SI{32.5}{\electronvolt}). The RECP-UCCSD(T)+SOC method is in good
agreement with ionization energies of Pa computed in
Ref.\nocite{cao:2003}\citenum{cao:2003} (deviations $<5\,\%$) and
experimental ionization energy and electron
affinity \cite{edlen:1969,blondel:2001} of F (deviations $<1\,\%$).  

From \emph{ab initio} calculations we find that the dissociation
channel \ce{Pa^3+} $+$ \ce{F+} is placed $\sim\SI{10}{eV}$ below the
equilibrium energy of \ce{PaF^4+}. A crude estimate for a repulsive
potential $V(r)$ as a function of the internuclear separation $r$ can
be obtained from purely repulsive Coulomb and Pauli
potentials \cite{harvey:2003,walker:2001}. Assuming that the ionic
radius $r_0$ is equal for charge separation and homolytic
dissociation, we model the Pauli repulsion by a Lennard-Jones
potential $C^{(12)}/(r+r_0)^{12}-C^{(6)}/(r+r_0)^{6}$ for a fixed
dissociation energy (indicated by crosses in 
Fig.~\ref{fig: setup_diss} 
on the right), which is fitted to the ground state
potentials of \ce{PaF^3+} and \ce{PaF^4+} computed on the level of
ZORA-cGKS-PBE0 as indicated by circles in 
Fig.~\ref{fig: setup_diss}
on the right to obtain $V(r)=\Delta E+q_1q_2/(4\pi\epsilon_0
(r+r_0))+C^{(12)}/(r+r_0)^{12}$. Here $\Delta E$ is received at the
level of ZORA-cGKS-PBE0 as described above and $q_1q_2$ is determined
by the charge of the fragments. The fit of the ground state potential
of \ce{PaF^3+} yields $r_0=\SI{1.55}{\angstrom}$,
$C^{(12)}=\SI{1.2e7}{\electronvolt\angstrom}^{12}$ and
$C^{(6)}=\SI{1.5e4}{\electronvolt\angstrom}^{6}$. The fit of the
ground state potential of \ce{PaF^4+} yields
$r_0=\SI{1.00}{\angstrom}$,
$C^{(12)}=\SI{6.9e5}{\electronvolt\angstrom}^{12}$ and
$C^{(6)}=\SI{2.7e3}{\electronvolt\angstrom}^{6}$.  Relative to the
ground state energy of \ce{PaF^3+} we receive for the \ce{Pa^2+} $+$
\ce{F+} channel $\Delta E=\SI{3.8}{\electronvolt}$,
$r_0=\SI{1.55}{\angstrom}$,
$C^{(12)}=\SI{1.2e7}{\electronvolt\angstrom}^{12}$,
$q_1q_2/(4\pi\epsilon_0)=\SI{28.8}{\electronvolt\angstrom}$ yielding
an avoided crossing at $>\SI{30}{\angstrom}$ and for the \ce{Pa^3+}
$+$ \ce{F+} channel $\Delta E=\SI{35.5}{\electronvolt}$,
$r_0=\SI{1.00}{\angstrom}$,
$C^{(12)}=\SI{6.9e5}{\electronvolt\angstrom}^{12}$,
$q_1q_2/(4\pi\epsilon_0)=\SI{43.2}{\electronvolt\angstrom}$ yielding
an avoided crossing at $>\SI{2.4}{\angstrom}$ (see 
Fig.~\ref{fig: setup_diss} 
right). This suggests that \ce{PaF^4+} could be
meta-stable with a dissociation barrier $>\SI{1}{\electronvolt}$.  In
this model the potential for the \ce{Pa^4+} $+$ \ce{F-} channel is
modeled as
$V(r)=\SI{32}{\electronvolt}-\SI{57.6}{\electronvolt\angstrom}/(r+\SI{1.55}{\angstrom})+\SI{1.2e7}{\electronvolt\angstrom}^{12}/(r+\SI{1.55}{\angstrom})^{12}-\SI{1.7e4}{\electronvolt\angstrom}^{6}/(r+\SI{1.55}{\angstrom})^{6}$.

In calculations of the dissociation energy (Table
\ref{tab:diss_channel}) the mean-field RECP-ROHF and ZORA-cGHF methods
underestimate the dissociation energy dramatically as missing electron
correlation destabilizes considerably the Fluoride ions (see also
Table \ref{tab:ei_ea}). Moreover, in the ZORA-cGHF method spin-orbit
coupling and spin-polarization are incorporated self-consistently. As
both effects stabilize the Pa cations, the ZORA-cGHF values are much
lower than those at correlated levels of theory due this imbalanced
consideration of electron-correlation effects.

\subsection*{Electronic excitation energies}
In calculations of the electronic spectra (Table
\ref{tab:spectroscopy}) ZORA-cGHF results are in a good agreement with
DC-FSCCSD results, with deviations in excitation wavenumbers of maximally
\SI{600}{\per\centi\metre} (for higher excitations
\SI{1100}{\per\centi\metre}) and deviations of harmonic vibrational wavenumbers
and equilibrium bond lengths  $\leq10\,\%$. We expect other molecular
properties at the level of ZORA-cGHF to be accurate within $10\,\%$.
Furthermore, the effect of a larger basis set (ANO-RCC vs dyall.3aez
vs dyall.4aez) in FSCCSD calculations is found to be $<5\,\%$. 

\subsection*{Visualisation of nuclear densities}
Nuclear charge densities with deformation of order $n$ visualised in
Fig.~\ref{fig: schiff_moment} 
on the bottom are realised as spherical
plots of a Rayleigh expansion with axial symmetry \(
R_n(\theta,\phi,r)= \left[1+\sum_{l=2}^n a_l r^l
Y_{l,0}(\theta,\phi)\right] \) averaged over the radial part as \(
\int R_n(\theta,\phi,r)\exp[-2/(3\Braket{r^2})r^2]r^2\mathrm{d}r\),
with $a_2=0.231$, $a_3=0.097$, $a_4=0.04$ and $\Braket{r^2}=1$. These
coefficients are chosen for optimal representation and have no
physical meaning. The Schiff moment operator can be written as
$\hat{S}=(r^3-5/2\Braket{r^2}r) Y_{10}(\theta,\phi)$. The
corresponding Schiff moments are modeled in a spherical plot of the
angular function \( S(\theta,\phi)=\int \hat{S} R_n(\theta,\phi,r)
\exp[-2/(3\Braket{r^2})r^2] r^2\mathrm{d}r \), with the resulting
moment being calculated as $\mathcal{S}=\int
S(\theta,\phi)\sin(\theta)\mathrm{d}\theta\mathrm{d}\phi$.

\clearpage

%%%%%%%%%%%%%%%%diss channels%%%%%%%%%%%%%%
\begin{table}
\small
\begin{threeparttable}
\caption{\textbf{Dissociation energies of \ce{PaF^{3+}} given for the three most
probable dissociation channels.} The dissociation channels correspond to 
$D_e(\mathrm{A}^{n+}+\mathrm{B}^{[+,0,-]})=E(\mathrm{A}^{n+})+E(\mathrm{B}^{[+,0,-]})-E(\mathrm{AB}^{[(n+1)+,n+,(n-1)+]})$.
}
\label{tab:diss_channel}
\begin{tabular}{l
S[round-mode=places,round-precision=1]
S[round-mode=places,round-precision=1]
S[round-mode=places,round-precision=1]}
\toprule
{Method} &  
{$D_e(\text{Pa}^{2+} $+$  \text{F}^+) /\text{eV}$} & 
{$D_e(\text{Pa}^{3+} $+$  \text{F})   /\text{eV}$} & 
{$D_e(\text{Pa}^{4+} $+$  \text{F}^-) /\text{eV}$} 
\\
\midrule
RECP-ROHF         & 2.71901   &  4.23592  &  30.8113 \\
ZORA-cGHF\tnote{*}& 0.3212    &  2.69404  &  30.5286 \\
RECP-UKS-B3LYP    & 4.738     &  5.64466  &  32.7052 \\
ZORA-cGKS-B3LYP   & 4.06224   &  4.73454  &  32.1977 \\
RECP-UKS-PBE0     & 4.8274    &  5.61606  &  32.3916 \\
ZORA-cGKS-PBE0    & 3.795     &  4.73436  &  32.4771 \\
RECP-UCCSD(T)	  & 5.11187   &  5.79902  &  32.0132 \\
RECP-UCCSD(T)+SOC & 4.07947   &  4.91732  &  32.0987 \\
\bottomrule
\end{tabular}
\begin{tablenotes}
\item[*] The ZORA-cGHF method underestimates the dissociation energy
dramatically (see discussion in the methods section).
\end{tablenotes}
\end{threeparttable}
\end{table}
%%%%%%%%%%%%%%%%TABLE END%%%%%%%%%%%%%%%%%%

\clearpage

%%%%%%%%%%%E_i and E_ea%%%%%%%%%%%%
\begin{threeparttable}
\small
\caption{\textbf{Relevant $n$-th ionization energies $E_{\text{i}}$ and
electron affinities $E_{\text{ea}}$ for Pa and F in comparison with
literature (lit) values.} Relative deviation is given as 
 $\text{dev} = (\text{calculation} -\text{lit})/\text{lit}$.}
\begin{tabular}{l
S[table-format=-2.1,round-mode=places,round-precision=1]
S[table-format=-1.0,round-mode=places,round-precision=0]
S[table-format=-2.1,round-mode=places,round-precision=1]
S[table-format=-1.0,round-mode=places,round-precision=0]
S[table-format=-2.1,round-mode=places,round-precision=1]
S[table-format=-1.0,round-mode=places,round-precision=0]
S[table-format=-1.2,round-mode=places,round-precision=2]
S[table-format=-1.0,round-mode=places,round-precision=0]
}
\toprule
{Method} &  
{$E_{\text{i}}\left( \text{Pa}^{2+} \right) /\text{eV}$} & 
{$\text{dev} \  / \ \si{\percent}$} &
{$E_{\text{i}}\left( \text{Pa}^{3+} \right) /\text{eV}$} & 
{$\text{dev} /\si{\percent}$} &
{$E_{\text{i}}\left( \text{F} \right) /\text{eV}$} & 
{$\text{dev} /\si{\percent}$} &
{$E_{\text{ea}}\left( \text{F} \right) /\text{eV}$} &
{$\text{dev} /\si{\percent}$}	\\
\midrule
RECP-ROHF         &  17.2217        & -7.9      &  27.8883     &  -9.74 &  15.7048  &  -9.742   &  1.31        &  -64.17	\\
ZORA-cGHF         &  18.0127        & -3.7      &  29.0992     &  -5.8  &  15.64    &  -10.1    &  1.265       &  -62.8  \\	
RECP-UKS-B3LYP    &  18.3612        & -1.548    &  30.2963     &  -1.985&  17.4547  &  0.183    &  3.23576     &  -4.83\\
ZORA-cGKS-B3LYP	  &  18.4366        & -1.4      &  31.13       &  0.71  &  17.76    &  1.94     &  3.67        &  7.94	\\		
RECP-UKS-PBE0     &  18.42          & -1.23     &  30.204      &  -2.28 &  17.6314  &  1.197    &  3.42846     &  0.837\\
ZORA-cGKS-PBE0	  &  18.43          & -1.44     &  31.1428     &  0.786 &  17.43    &  0.06     &  3.400       &  0.0	\\
RECP-UCCSD(T)     &  18.037         & -3.5      &  29.59       &  -4.2  &  17.35    &  -0.2     &  3.4         &  0.00	\\
RECP-UCCSD(T)+SOC &  18.047         & -3.2      &  30.5288     &  -1.2  &  17.35    &  -0.2     &  3.4         &  0.00	\\
Literature        &  18.65~\tnote{a} & {\textemdash}& 30.91~\tnote{a}
 &{\textemdash}& 17.42282~\tnote{b} &{\textemdash} &3.40~\tnote{b}&{\textemdash}\\
\bottomrule
\end{tabular}
\label{tab:ei_ea}
\begin{tablenotes}
\item[a] Scalar relativistic effective
core potential calculations at the level of complete active space self-consistent field
(RECP-CASSCF) with a correction for
spin-orbit coupling by comparison to multi-configuration
Dirac-Hartree-Fock (MCDHF) calculations.\cite{cao:2003}
\item[b]Experimental data. $E_{\text{i}}\left( \text{F} \right) /
\text{eV}$ was measured in Ref.\nocite{edlen:1969}\citenum{edlen:1969} and
$E_{\text{ea}}\left( \text{F} \right) / \text{eV}$ was measured in
Ref.\nocite{blondel:2001}\citenum{blondel:2001}.
\end{tablenotes}
\end{threeparttable}
%%%%%%%%%%%%%%%%TABLE END%%%%%%%%%%

\clearpage

%%%%%%%%%%%%%PaF^4+%%%%%%%%%%%%%%
\begin{table}
\small
\begin{threeparttable}
\caption{\textbf{Dissociation energies for \ce{PaF^{4+}} given for the
two most probable dissociation channels as
$D_\mathrm{e}(\text{channel})$.}} 
\begin{tabular}{l
S[round-mode=places,round-precision=1,table-column-width=2.8cm]
S[round-mode=places,round-precision=1,table-column-width=2.8cm]}
\toprule
{Method} &  
{$D_\mathrm{e}(\text{Pa}^{4+} $+$ \text{F}) /\text{eV}$} & 
{$D_\mathrm{e}(\text{Pa}^{3+} $+$ \text{F}^{+}) /\text{eV}$}\\ 
\midrule
RECP-ROHF         & 0.56489   & -11.6187  \\
ZORA-cGHF         & 0.0924723 & -13.3669   \\
RECP-UKS-B3LYP    & 3.47459   &  -9.367   \\
ZORA-cGKS-B3LYP   & 3.1540    & -10.0935   \\
RECP-UKS-PBE0     & 3.41546   & -9.15719   \\
ZORA-cGKS-PBE0    & 3.30188   & -10.3485  \\
RECP-UCCSD(T)     & 3.65527   & -8.5897    \\
RECP-UCCSD(T)+SOC & 3.6       & -9.7     \\ 
\bottomrule
\end{tabular}
\label{tab:paf4+}
\end{threeparttable}
\end{table}

\clearpage

\begin{table}
\scriptsize
\begin{threeparttable}
\caption{\small \textbf{Spectroscopically relevant properties of the eight energetically
lowest electronic states of \ce{PaF3^+}.} Equilibrium bond length
$r_\mathrm{e}$, harmonic vibrational wavenumber $\tilde{\omega}_\mathrm{e}$ and
excitation wavenumber $\tilde{T}_\mathrm{e}$ estimated as vertical excitation
energy are shown at the level of Dirac--Coulomb Fock-Space Coupled Cluster
(DC-FSCCSD) with two different basis sets and at the level of Zeroth Order
Regular Approximation complex Generalized Hartree-Fock (ZORA-cGHF). The
projection of the electronic orbital angular momentum quantum number on the
molecular axis $\Lambda$, the squared transition dipole moment $\left|
\vec{\mu} \right| ^2 $ and hyperfine coupling constant along the molecular axis
$A_{\parallel}$ are given. Hyperfine
coupling constants were calculated using $\mu \left( ^{231}\text{Pa} \right)=
2.01~\mu_{\text{N}}$ and $I=3/2$ \cite{axe:1961}. The DC-FSCCSD/dyall.ae4z
results were computed at the equilibrium bond length taken from the
ZORA-cGKS-B3LYP calculations.} \label{tab:spectroscopy}
\begin{tabular}{l
l
S[table-format=1.2,round-mode=figures,round-precision=3 ]
S[table-format=3.0,round-mode=figures,round-precision=3 ]
S[table-format=4.0,round-mode=figures,round-precision=3 ]
S[table-format=1.1,round-mode=places,round-precision=1  ]
S[table-format=1e-1,round-mode=figures,round-precision=1]
S[table-format=4.0,round-mode=figures,round-precision=3 ]}
\toprule
{State} &
{Method} &
{$r_\mathrm{e} /\si{\angstrom}$} &
{$\tilde{\omega}_\mathrm{e} /\si{\per\centi\metre}$} &
{$\tilde{T}_\mathrm{e} /\si{\per\centi\metre}$} &
{$\Lambda$} &
{$\left| \vec{\mu}\right|^2 / (e^2\si{\bohr^2})$} &
{$A_{\parallel} /\si{\mega\hertz}$}\\
\midrule
\multirow{5}{*}{$(X)5/2$}& ZORA-cGHF & 1.87353 &  846.380 & {\textemdash} & 2.90066 & & -983  \\
& ZORA-cGKS-B3LYP     & 1.88814 & 828.172 & {\textemdash} & 2.79401 &  & -1020 \\
& ZORA-cGKS-PBE0      & 1.86795 & 827.725 & {\textemdash} & 2.79999 &  & -1023 \\
& DC-FSCCSD/dyall.ae3z&1.84947 & 859.181 & {\textemdash} & \\
& DC-FSCCSD/ANO-RCC  & 1.84791 & 891.877 & {\textemdash} & \\
\\
\multirow{5}{*}{$(1){3/2}$}& ZORA-cGHF & 1.86537 & 829.242 & 1254.127 & 1.82234 & 3.3E-3 & -953 \\
& ZORA-cGKS-B3LYP     & 1.88163 & 816.174 & 152.009 & 1.87812  & 4.11 E-3 & -1138\\
& ZORA-cGKS-PBE0      & 1.86086 & 814.690 & 185.456 & 1.87399 & 3.73 E-3 & -1133 \\
& DC-FSCCSD/dyall.ae3z&1.84947 & 844.064 & 657.636 & \\
& DC-FSCCSD/dyall.ae4z & & & 751.657&\\
& DC-FSCCSD/ANO-RCC & 1.84182 & 876.311 & 646.375 & \\
\\
\multirow{5}{*}{$(1)1/2$}& ZORA-cGHF & 1.87351 & 839.788 & 3023.307 & 0.62204 & 6.0E-8 & -1054\\
& ZORA-cGKS-B3LYP     & 1.89139 & 831.242 & 3465.73155149 & 0.79396 & 2.19 E-9 & -1598\\
& ZORA-cGKS-PBE0      & 1.8714  & 830.029 & 3535.739 & 0.78303 & 1.49 E-11 & -1556 \\
& DC-FSCCSD/dyall.ae3z&1.85508 & 852.469 & 3060.091 & \\
& DC-FSCCSD/dyall.ae4z & & & 3051.224 &\\
& DC-FSCCSD/ANO-RCC & 1.85263  & 885.230 & 2991.7663&\\
\\
\multirow{3}{*}{$(1)7/2$}& ZORA-cGHF & 1.86506  & 849.177 & 5523.92 & 2.99972 & 1.7E-4 & -328 \\
& DC-FSCCSD/dyall.ae3z& 1.8502   & 861.519& 5540.48  \\
& DC-FSCCSD/dyall.ae4z & & & 5550.1639& \\
& DC-FSCCSD/ANO-RCC   & 1.84863  & 894.510& 5551.8561       \\
\\
\multirow{3}{*}{$(1)5/2$}& ZORA-cGHF & 1.86506  & 831.077  & 6444 & 2.11498 & 1.5E-3 & -329 \\
& DC-FSCCSD/dyall.ae3z& 1.84388  & 846.037& 5787.3327 \\
& DC-FSCCSD/dyall.ae4z & & & 5680.520 & \\
& DC-FSCCSD/ANO-RCC &   1.8402   & 878.201& 5677.9909\\
\\
\multirow{3}{*}{$(2)3/2$}& ZORA-cGHF & 1.87135  & 838.147  & 8001.94 & 1.17057 & 5.3E-4 & -329\\
& DC-FSCCSD/dyall.ae3z& 1.85045  & 853.019& 7809.933\\
& DC-FSCCSD/dyall.ae4z & & & 7827.132  & \\
& DC-FSCCSD/ANO-RCC &   1.8484   & 884.991& 7784.2098\\
\\
\multirow{3}{*}{$(2)1/2$}& ZORA-cGHF & 1.87388  & 840.915  & 8893  & 0.39891 & 1.4E-11  & -473\\
& DC-FSCCSD/dyall.ae3z& 1.8560  &857.4997 & 8941.502\\
& DC-FSCCSD/dyall.ae4z & & & 9036.379  &\\
& DC-FSCCSD/ANO-RCC &   1.85337  &889.782  & 8964.653 \\
\\
\multirow{3}{*}{$(3)3/2$}& ZORA-cGHF & 1.83692  & 887.505  & 30022.796 & 1.98519 & 0.53 & -2600\\
& DC-FSCCSD-dyall-ae3z& 1.82440  & 882 & 29166.40 \\
& DC-FSCCSD/dyall.ae4z & & & 29890.9879&\\
& DC-FSCCSD-ANO-RCC &   1.824 & 901 & 28863.1   \\
\bottomrule
\end{tabular}
\end{threeparttable}
\end{table}

\clearpage

%%%%%%%%%%%%%%%Transition properties%%%%%%%%%%

%{
%\setlength\extrarowheight{5pt}
\begin{table}
\scriptsize
\begin{threeparttable}
\caption{\textbf{The transition properties relevant for investigating
laser-cooling are shown for the transitions between the eight lowest states of
\ce{PaF^{3+}}}. Excitation energies $T_e$, Einstein coefficients $A$ for
spontaneous emission from electronic states (as detailed in the methods
section) and Franck--Condon factors for the 0-0 transition ($f^{(0)}$) and
cumulated for the 1-0 transition ($f^{(1)}$) are computed on the level of
ZORA-cGHF. The life-time of electronic states $\tau_e$ is estimated from the
electronic Einstein coefficients as $(\sum_{a}^i A_i^a)^{-1}$ for each state
$i$, where $A_i^a$ is the Einstein coefficient for the electronic spontaneous
emission from $i$ to $a$. Franck--Condon factors in parantheses indicate that
the corresponding vibrational level lies above the vibrational ground state of
the next higher electronic state.}
\begin{tabular}{ l l|
S[group-digits=false,table-text-alignment=right,input-symbols= (),table-format=5.0e+1]| 
S[group-digits=false,table-text-alignment=right,input-symbols= (),table-format=5.0e+1]|
S[group-digits=false,table-text-alignment=right,input-symbols= (),table-format=5.0e+1]|
S[group-digits=false,table-text-alignment=right,input-symbols= (),table-format=5.0e+1]| 
S[group-digits=false,table-text-alignment=right,input-symbols= (),table-format=5.0e+1]| 
S[group-digits=false,table-text-alignment=right,input-symbols= (),table-format=5.0e+1]| 
S[group-digits=false,table-text-alignment=right,input-symbols= (),table-format=5.0e+1] 
}
\toprule
                       & &{$(1)3/2$}&{$(1)1/2$} & {$(1)7/2$}& {$(1)5/2$}& {$(2)3/2$}& {$(2)1/2$}& {$(3)3/2$}\\ 
\midrule
&{$\tau_e / \si{s}$}  
& 7e-2       & 6E-3       & 2E-2      & 7E-4     & 4E-4    & 6E-6  & 2E-8 \\    
\midrule
\multirow5{*}{$(X)5/2$}& {$T_e/\si{cm^{-1}}$}                                                              & 1300       & 3000     & 5500    & 6400    & 8000    & 8800  & 30000 \\
                         & {$T_e /\si{nm}$}                                                                 & 7700       & 3300     & 1800    & 1600    & 1300    & 1100  & 300 \\
                         & {$A /\si{s^{-1}}$}                                                               & 1e1        & 3E-3       & 5e1      & 1E3        & 5E2       & 2E-5    & 3E7 \\
                         & {$f^{(0)}$}                                                                             & 0.9855     &0.9856    & 0.9994  & 0.9844  & 0.9989  &0.9999 & 0.7391 \\
			 & {$f^{(1)}$}                                                                             & (0.9999)   & 0.9999  & 0.9999 &0.9999 & 0.9999  & 0.9999 & 0.9570 \\
\hline
\multirow5{*}{$(1)3/2$}& {$T_e  /\si{cm^{-1}}$}                                                              &              & 1800    & 4300    & 5200     & 6700    & 7500    & 29000\\
                         & {$T_e/\si{nm}$}                                                                   &              & 5500    & 2300    & 1900     & 1500    & 1300    & 300\\
                         & {$A  /\si{s^{-1}}$}                                                                &              & 2E2       & 2E-5      & 3e2      & 2E3    & 1E5       & 1E7\\
			 & {$f^{(0)}$}                                                                             &       & 0.9857  & 0.9795  & 0.9999  & 0.9922 & 0.9843  & 0.8342\\
			 & {$f^{(1)}$}                                                                             &       & 0.9997  & 0.9994 & 0.9999 &0.9999 & 0.9997  & 0.9798 \\
\hline
\multirow5{*}{$(1)1/2$}& {$T_e  /\si{cm^{-1}}$}                                                              &       &        & 2500    & 3400    & 5000    & 5800    & 27000   \\
                         & {$T_e/\si{nm}$}                                                                   &       &        & 4000    & 2900    & 2000    & 1700    & 370   \\
                         & {$A  /\si{s^{-1}}$}                                                               &       &        & 5E-7      & 3E-4      & 1e0       & 1E4       & 1E6\\
			 & {$f^{(0)}$}                                                                             &       &        & 0.9994  & 0.9846  & 0.9989 & 0.9969  & 0.7399 \\
			 & {$f^{(1)}$}                                                                             &       &        & 0.9999 & 0.9999 & 0.9999 & 0.9999  & 0.9563 \\
\hline
\multirow5{*}{$(1)7/2$}& {$T_e  /\si{cm^{-1}}$}                                                              &       &        &        & 920     & 2500    & 3300    & 24000\\
                         & {$T_e/\si{nm}$}                                                                   &       &        &        & 10900   & 4000    & 3030    & 400\\
                         & {$A  /\si{s^{-1}}$}                                                                 &       &        &        & 3       & 3E-3      & 2E-6      & 9E-2 \\ 
			 & {$f^{(0)}$}                                                                             &       &        &        & 0.9782  & 0.9969 & 0.9997  & 0.7194 \\
			 & {$f^{(1)}$}                                                                             &       &        &        & (0.9999) & 0.9999 & 0.9999  & 0.9509 \\
\hline
\multirow5{*}{$(1)5/2$}& {$T_e  /\si{cm^{-1}}$}                                                              &       &        &        &        & 1600    & 2400 & 24000\\
                         & {$T_e/\si{nm}$}                                                                   &       &        &        &        & 6300    & 4200 & 400\\
                         & {$A  /\si{s^{-1}}$}                                                                 &       &        &        &        & 1E2       & 2E-6   & 1E6 \\
			 & {$f^{(0)}$}                                                                             &       &        &        &        & 0.9914 & 0.9831  & 0.8374 \\
                         & {$f^{(1)}$}                                                                             &       &        &        &        & (0.9999)& 0.9997   & 0.9806 \\
\hline
\multirow5{*}{$(2)3/2$}& {$T_e  /\si{cm^{-1}}$}                                                              &       &        &        &        &        & 820  & 22000\\
                         & {$T_e/\si{nm}$}                                                                   &       &        &        &        &        & 12200& 500\\
                         & {$A  /\si{s^{-1}}$}                                                               &       &        &        &        &        & 1e1   & 3E5 \\
			 & {$f^{(0)}$}                                                                             &       &        &        &        &        & (0.9986)  & 0.7768 \\
                         & {$f^{(1)}$}                                                                             &       &        &        &        &        & (0.9999)  & 0.9670\\
\hline
\multirow5{*}{$(2)1/2$}& {$T_e  /\si{cm^{-1}}$}                                                              &       &        &        &        &        &        & 21000 \\
                         & {$T_e/\si{nm}$}                                                                   &       &        &        &        &        &        & 500 \\
                         & {$A  /\si{s^{-1}}$}                                                                 &       &        &        &        &        &        & 1E3 \\
			 & {$f^{(0)}$}                                                                             &       &        &        &        &        &        & 0.7353 \\
                         & {$f^{(1)}$}                                                                             &       &        &        &        &        &        & 0.9550 \\
\bottomrule
\end{tabular}
\label{tab:S7}
\end{threeparttable}
\end{table}
%}

\clearpage

%%%%%%%%%%%%%%%%%Properties%%%%%%%%%%%%%%
\begin{threeparttable}
\small
\caption{\textbf{$\mathcal{P}$,$\mathcal{T}$-odd electronic structure
parameters for PaF$^{3+}$ are shown.} The properties were
calculated on the level of ZORA-cGHF and ZORA-cGKS-B3LYP and PBE0. For
computation of the magnetic interaction with the proton EDM
$W_\mathrm{m}$ $\mu \left( ^{231}\text{Pa} \right)\simeq 2.01
\mu_{\text{N}}$ \cite{axe:1961} and $I=3/2$ were used.  }
\begin{tabular}{l
l
S[table-format=-5.0,round-mode=figures,round-precision=2]
S[table-format=-1.2,round-mode=figures,round-precision=2]
S[table-format=-4.0,round-mode=figures,round-precision=2]
S[table-format=-1.1,round-mode=figures,round-precision=2]
S[table-format=-2.1,round-mode=figures,round-precision=2]
S[table-format=-0.4,round-mode=figures,round-precision=2]
}
\toprule
{State} &
{Method} &
{$W_{\mathcal{S}} / \frac{e}{4\pi\epsilon_0a_0^4}  $} &
{$W_{\text{m}} / \frac{10^{18}\,h\,\si{Hz}}{e\,\si{cm}}  $} &
{$W_{\text{T}} / (h\,\si{Hz})  $} &
{$W_{\text{d}} / \frac{10^{24}\,h\,\si{Hz}}{e\,\si{cm}}  $} &
{$W_{\text{s}} /(h\,\si{kHz}) $} & 
{$W_{\mathcal{M}} / \frac{10^{33}\,h\,\si{Hz}}{c\,e\,\si{cm^2}}  $} 
\\
\midrule
\multirow{3}{*}{$X5/2$}& ZORA-cGHF     & -72249 & 6.3219 & -6727 & 0.655  & 4.155     & 0.0378    \\ 
& ZORA-cGKS-B3LYP 	                  & -57653 & 4.9498 & -5300 & 0.411  & 2.530     & 0.0228    \\ 
& ZORA-cGKS-PBE0  	                  & -58276 & 4.8907 & -5384 & 0.411  & 2.870     & 0.0269    \\ 
\\                                                                        
\multirow{3}{*}{$(1){3/2}$}& ZORA-cGHF & -65649 & 5.5487 & -6182 & 1.391  & 9.342     & -0.02231    \\ 
& ZORA-cGKS-B3LYP 	                  & -53402 & 4.5052 & -4949 & 0.918  & 5.777     & -0.00225    \\   
& ZORA-cGKS-PBE0  	                  & -53836 & 4.4201 & -5017 & 1.068  & 6.781     & -0.00175    \\ 
\\                                                                        
\multirow{3}{*}{$(1)1/2$}& ZORA-cGHF   & -58732 & 4.8496 & -5588 & 2.916  & 19.561    & -0.1033     \\ 
& ZORA-cGKS-B3LYP  	                  & -48716 & 4.0637 & -4803 & 1.789  & 11.860    & -0.1254     \\ 
& ZORA-cGKS-PBE0  	                  & -48854 & 3.9480 & -4622 & 2.016  & 13.349    & -0.1413     \\ 
\\                                                                        
$(1)7/2$ & ZORA-cGHF                   & -72966 & 6.3830 & -6782 & -0.479 & -2.960    & -0.0303     \\ 
$(1)5/2$ & ZORA-cGHF                   & -67525 & 5.7473 & -6353 & -0.283 & -1.880    & -0.00144    \\ 
$(2)3/2$ & ZORA-cGHF                   & -60781 & 5.1279 & -5543 & -0.220 & -1.562    & 0.0609    \\ 
$(2)1/2$ & ZORA-cGHF                   & -57692 & 4.8849 & -5455 & -0.361 & -2.599    & 0.1355    \\ 
$(3)3/2$ & ZORA-cGHF                   & -70043 & 5.8361 & -6566 & 3.17   & 18.510    & 0.1386    \\ 
\bottomrule
\end{tabular}
\label{tab:properties}
\end{threeparttable}
%%%%%%%%%%%%%%TABLE END%%%%%%%%%%%%%
\clearpage

\begin{thebibliography}{10}
\expandafter\ifx\csname url\endcsname\relax
  \def\url#1{\texttt{#1}}\fi
\expandafter\ifx\csname urlprefix\endcsname\relax\def\urlprefix{URL }\fi
\providecommand{\bibinfo}[2]{#2}
\providecommand{\eprint}[2][]{\url{#2}}

\bibitem{spergel:2015}
\bibinfo{author}{Spergel, D.~N.}
\newblock \bibinfo{title}{The dark side of cosmology: Dark matter and dark
  energy}.
\newblock \emph{\bibinfo{journal}{Science}} \textbf{\bibinfo{volume}{347}},
  \bibinfo{pages}{1100--1102} (\bibinfo{year}{2015}).
\newblock \urlprefix\url{https://science.sciencemag.org/content/347/6226/1100}.

\bibitem{canetti:2012}
\bibinfo{author}{Canetti, L.}, \bibinfo{author}{Drewes, M.} \&
  \bibinfo{author}{Shaposhnikov, M.}
\newblock \bibinfo{title}{Matter and antimatter in the universe}.
\newblock \emph{\bibinfo{journal}{New J. Phys.}} \textbf{\bibinfo{volume}{14}},
  \bibinfo{pages}{095012} (\bibinfo{year}{2012}).
\newblock
  \urlprefix\url{https://doi.org/10.1088%2F1367-2630%2F14%2F9%2F095012}.

\bibitem{gross:1996}
\bibinfo{author}{Gross, D.~J.}
\newblock \bibinfo{title}{{The role of symmetry in fundamental physics}}.
\newblock \emph{\bibinfo{journal}{Proc. Natl. Acad. Sci. USA}}
  \textbf{\bibinfo{volume}{93}}, \bibinfo{pages}{14256--14259}
  (\bibinfo{year}{1996}).
\newblock \urlprefix\url{http://www.pnas.org/content/93/25/14256.full}.

\bibitem{andreev:2018}
\bibinfo{author}{Andreev, V.} \emph{et~al.}
\newblock \bibinfo{title}{Improved limit on the electric dipole moment of the
  electron}.
\newblock \emph{\bibinfo{journal}{Nature}} \textbf{\bibinfo{volume}{562}},
  \bibinfo{pages}{355--360} (\bibinfo{year}{2018}).
\newblock \urlprefix\url{https://doi.org/10.1038/s41586-018-0599-8}.

\bibitem{demille:2015}
\bibinfo{author}{{DeMille}, D.}
\newblock \bibinfo{title}{Diatomic molecules, a window onto fundamental
  physics}.
\newblock \emph{\bibinfo{journal}{Physics Today}}
  \textbf{\bibinfo{volume}{68}}, \bibinfo{pages}{34--40}
  (\bibinfo{year}{2015}).
\newblock \urlprefix\url{https://doi.org/10.1063/PT.3.3020}.
\newblock \eprint{https://doi.org/10.1063/PT.3.3020}.

\bibitem{kozlov:2018}
\bibinfo{author}{Kozlov, M.~G.}, \bibinfo{author}{Safronova, M.~S.},
  \bibinfo{author}{Crespo L{\'o}pez-Urrutia, J.~R.} \&
  \bibinfo{author}{Schmidt, P.~O.}
\newblock \bibinfo{title}{Highly charged ions: Optical clocks and applications
  in fundamental physics}.
\newblock \emph{\bibinfo{journal}{Rev. Mod. Phys.}}
  \textbf{\bibinfo{volume}{90}}, \bibinfo{pages}{045005}
  (\bibinfo{year}{2018}).
\newblock
  \urlprefix\url{https://link.aps.org/doi/10.1103/RevModPhys.90.045005}.

\bibitem{uzan:2003}
\bibinfo{author}{Uzan, J.-P.}
\newblock \bibinfo{title}{The fundamental constants and their variation:
  observational and theoretical status}.
\newblock \emph{\bibinfo{journal}{Rev. Mod. Phys.}}
  \textbf{\bibinfo{volume}{75}}, \bibinfo{pages}{403--455}
  (\bibinfo{year}{2003}).
\newblock \urlprefix\url{https://link.aps.org/doi/10.1103/RevModPhys.75.403}.

\bibitem{berengut:2010}
\bibinfo{author}{Berengut, J.~C.}, \bibinfo{author}{Dzuba, V.~A.} \&
  \bibinfo{author}{Flambaum, V.~V.}
\newblock \bibinfo{title}{Enhanced laboratory sensitivity to variation of the
  fine-structure constant using highly charged ions}.
\newblock \emph{\bibinfo{journal}{Phys. Rev. Lett.}}
  \textbf{\bibinfo{volume}{105}}, \bibinfo{pages}{120801}
  (\bibinfo{year}{2010}).
\newblock
  \urlprefix\url{https://link.aps.org/doi/10.1103/PhysRevLett.105.120801}.

\bibitem{drewsen:1998}
\bibinfo{author}{Drewsen, M.}, \bibinfo{author}{Brodersen, C.},
  \bibinfo{author}{{Hornek\ae{}r}, L.}, \bibinfo{author}{Hangst, J.~S.} \&
  \bibinfo{author}{Schifffer, J.~P.}
\newblock \bibinfo{title}{Large ion crystals in a linear paul trap}.
\newblock \emph{\bibinfo{journal}{Phys. Rev. Lett.}}
  \textbf{\bibinfo{volume}{81}}, \bibinfo{pages}{2878--2881}
  (\bibinfo{year}{1998}).
\newblock \urlprefix\url{https://link.aps.org/doi/10.1103/PhysRevLett.81.2878}.

\bibitem{bowe:1999}
\bibinfo{author}{Bowe, P.} \emph{et~al.}
\newblock \bibinfo{title}{Sympathetic crystallization of trapped ions}.
\newblock \emph{\bibinfo{journal}{Phys. Rev. Lett.}}
  \textbf{\bibinfo{volume}{82}}, \bibinfo{pages}{2071--2074}
  (\bibinfo{year}{1999}).
\newblock \urlprefix\url{https://link.aps.org/doi/10.1103/PhysRevLett.82.2071}.

\bibitem{nguyen:2011}
\bibinfo{author}{Nguyen, J. H.~V.} \emph{et~al.}
\newblock \bibinfo{title}{Challenges of laser-cooling molecular ions}.
\newblock \emph{\bibinfo{journal}{New. J. Phys.}}
  \textbf{\bibinfo{volume}{13}}, \bibinfo{pages}{063023}
  (\bibinfo{year}{2011}).
\newblock \urlprefix\url{https://doi.org/10.1088/1367-2630/13/6/063023}.

\bibitem{ivanov:2020}
\bibinfo{author}{Ivanov, M.~V.}, \bibinfo{author}{Jagau, T.-C.},
  \bibinfo{author}{Zhu, G.-Z.}, \bibinfo{author}{Hudson, E.~R.} \&
  \bibinfo{author}{Krylov, A.~I.}
\newblock \bibinfo{title}{In search of molecular ions for optical cycling: a
  difficult road}.
\newblock \emph{\bibinfo{journal}{Phys. Chem. Chem. Phys.}}
  \textbf{\bibinfo{volume}{22}}, \bibinfo{pages}{17075--17090}
  (\bibinfo{year}{2020}).
\newblock \urlprefix\url{http://dx.doi.org/10.1039/D0CP02921A}.

\bibitem{schroder:1999}
\bibinfo{author}{Schr\"oder, D.} \& \bibinfo{author}{Schwarz, H.}
\newblock \bibinfo{title}{Generation, stability, and reactivity of small,
  multiply charged ions in the gas phase}.
\newblock \emph{\bibinfo{journal}{J. Phys. Chem. A}}
  \textbf{\bibinfo{volume}{103}}, \bibinfo{pages}{7385--7394}
  (\bibinfo{year}{1999}).
\newblock \urlprefix\url{https://doi.org/10.1021/jp991332x}.

\bibitem{franzreb:2004}
\bibinfo{author}{Franzreb, K.} \emph{et~al.}
\newblock \bibinfo{title}{Gas-phase diatomic trications of {Se$_2^{3+}$},
  te{$_2^{3+}$}, and {LaF$^{3+}$}}.
\newblock \emph{\bibinfo{journal}{J. Chem. Phys.}}
  \textbf{\bibinfo{volume}{121}}, \bibinfo{pages}{12293--12302}
  (\bibinfo{year}{2004}).
\newblock \urlprefix\url{https://aip.scitation.org/doi/abs/10.1063/1.1821496}.

\bibitem{schroder:1999a}
\bibinfo{author}{Schr{\"o}der, D.}, \bibinfo{author}{Diefenbach, M.},
  \bibinfo{author}{Klap{\"o}tke, T.~M.} \& \bibinfo{author}{Schwarz, H.}
\newblock \bibinfo{title}{{UF$^{3+}$} -- a thermochemically stable diatomic
  trication with a covalent bond}.
\newblock \emph{\bibinfo{journal}{Angew. Chem. Int. Ed.}}
  \textbf{\bibinfo{volume}{38}}, \bibinfo{pages}{137--140}
  (\bibinfo{year}{1999}).
\newblock
  \urlprefix\url{https://onlinelibrary.wiley.com/doi/abs/10.1002/%28SICI%291521-3773%2819990115%2938%3A1/2%3C137%3A%3AAID-ANIE137%3E3.0.CO%3B2-M}.

\bibitem{khriplovich:1997}
\bibinfo{author}{Khriplovich, I.~B.} \& \bibinfo{author}{Lamoreaux, S.~K.}
\newblock \emph{\bibinfo{title}{\emph{CP} Violation without Strangeness}}
  (\bibinfo{publisher}{Springer}, \bibinfo{address}{Berlin},
  \bibinfo{year}{1997}).

\bibitem{chupp:2019}
\bibinfo{author}{Chupp, T.~E.}, \bibinfo{author}{Fierlinger, P.},
  \bibinfo{author}{Ramsey-Musolf, M.~J.} \& \bibinfo{author}{Singh, J.~T.}
\newblock \bibinfo{title}{Electric dipole moments of atoms, molecules, nuclei,
  and particles}.
\newblock \emph{\bibinfo{journal}{Rev. Mod. Phys.}}
  \textbf{\bibinfo{volume}{91}}, \bibinfo{pages}{015001}
  (\bibinfo{year}{2019}).
\newblock
  \urlprefix\url{https://link.aps.org/doi/10.1103/RevModPhys.91.015001}.

\bibitem{bates:1955}
\bibinfo{author}{Bates, D.~R.} \& \bibinfo{author}{Carson, T.~R.}
\newblock \bibinfo{title}{Doubly charged diatomic molecular ions}.
\newblock \emph{\bibinfo{journal}{Proceedings of the Physical Society. Section
  A}} \textbf{\bibinfo{volume}{68}}, \bibinfo{pages}{1199--1202}
  (\bibinfo{year}{1955}).
\newblock \urlprefix\url{https://doi.org/10.1088/0370-1298/68/12/417}.

\bibitem{migdalek:2007}
\bibinfo{author}{Migdalek, J.} \& \bibinfo{author}{Glowacz-Proszkiewicz, A.}
\newblock \bibinfo{title}{{Dirac--Fock} {$+$} core-polarization calculations of
  {E1} transitions in the francium isoelectronic sequence}.
\newblock \emph{\bibinfo{journal}{J. Phys. B}} \textbf{\bibinfo{volume}{40}},
  \bibinfo{pages}{4143--4154} (\bibinfo{year}{2007}).
\newblock \urlprefix\url{https://doi.org/10.1088/0953-4075/40/21/002}.

\bibitem{wyart:1981}
\bibinfo{author}{Wyart, J.-F.} \& \bibinfo{author}{Kaufman, V.}
\newblock \bibinfo{title}{Extended analysis of doubly ionized thorium ({Th}
  {III})}.
\newblock \emph{\bibinfo{journal}{Phys. Scr.}} \textbf{\bibinfo{volume}{24}},
  \bibinfo{pages}{941--952} (\bibinfo{year}{1981}).
\newblock \urlprefix\url{https://doi.org/10.1088/0031-8949/24/6/006}.

\bibitem{cao:2003}
\bibinfo{author}{Cao, X.} \& \bibinfo{author}{Dolg, M.}
\newblock \bibinfo{title}{Theoretical prediction of the second to fourth
  actinide ionization potentials}.
\newblock \emph{\bibinfo{journal}{Mol. Phys.}} \textbf{\bibinfo{volume}{101}},
  \bibinfo{pages}{961--969} (\bibinfo{year}{2003}).
\newblock \urlprefix\url{https://doi.org/10.1080/0026897021000046807}.
\newblock \eprint{https://doi.org/10.1080/0026897021000046807}.

\bibitem{edlen:1969}
\bibinfo{author}{Edl{\'e}n, B.}
\newblock \bibinfo{title}{On the identification of {Ar} x and {Ar} xiv in the
  solar corona and the origin of the unidentified coronal lines}.
\newblock \emph{\bibinfo{journal}{Sol. Phys.}} \textbf{\bibinfo{volume}{9}},
  \bibinfo{pages}{439--445} (\bibinfo{year}{1969}).
\newblock \urlprefix\url{https://doi.org/10.1007/BF02391668}.

\bibitem{eriksson:1963}
\bibinfo{author}{Eriksson, K. B.~S.} \& \bibinfo{author}{Isberg, H. B.~S.}
\newblock \bibinfo{title}{New measurements in the spectrum of atomic oxygen,
  {O} {I}}.
\newblock \emph{\bibinfo{journal}{Arkiv. Fysik}} \textbf{\bibinfo{volume}{24}},
  \bibinfo{pages}{549} (\bibinfo{year}{1963}).

\bibitem{eriksson:1971}
\bibinfo{author}{Eriksson, K. B.~S.} \& \bibinfo{author}{Pettersson, J.~E.}
\newblock \bibinfo{title}{New measurements in the spectrum of the neutral
  nitrogen atom}.
\newblock \emph{\bibinfo{journal}{Phys. Scr.}} \textbf{\bibinfo{volume}{3}},
  \bibinfo{pages}{211--217} (\bibinfo{year}{1971}).
\newblock \urlprefix\url{https://doi.org/10.1088/0031-8949/3/5/003}.

\bibitem{koch:1987}
\bibinfo{author}{Koch, W.} \& \bibinfo{author}{Frenking, G.}
\newblock \bibinfo{title}{Theoretical investigations of small multiply charged
  cations. ii. {CNe}$^{n+}$ ($1 \le n \le 4$)}.
\newblock \emph{\bibinfo{journal}{J. Chem. Phys.}}
  \textbf{\bibinfo{volume}{86}}, \bibinfo{pages}{5617--5624}
  (\bibinfo{year}{1987}).
\newblock \urlprefix\url{https://doi.org/10.1063/1.452538}.

\bibitem{isaev:2010}
\bibinfo{author}{Isaev, T.~A.}, \bibinfo{author}{Hoekstra, S.} \&
  \bibinfo{author}{Berger, R.}
\newblock \bibinfo{title}{Laser-cooled {RaF} as a promising candidate to
  measure molecular parity violation}.
\newblock \emph{\bibinfo{journal}{Phys. Rev. A}} \textbf{\bibinfo{volume}{82}},
  \bibinfo{pages}{052521} (\bibinfo{year}{2010}).

\bibitem{isaev:2013}
\bibinfo{author}{{Isaev}, T.~A.} \& \bibinfo{author}{{Berger}, R.}
\newblock \bibinfo{title}{Lasercooled radium monofluoride: A molecular
  all-in-one probe for new physics}.
\newblock \emph{\bibinfo{journal}{ArXiv e-prints}}
  \textbf{\bibinfo{volume}{1302.5682}}, \bibinfo{pages}{physics.chem--ph}
  (\bibinfo{year}{2013}).
\newblock \urlprefix\url{http://arxiv.org/abs/1302.5682}.
\newblock \eprint{1302.5682}.

\bibitem{garciaruiz:2020}
\bibinfo{author}{Garcia~Ruiz, R.~F.} \emph{et~al.}
\newblock \bibinfo{title}{Spectroscopy of short-lived radioactive molecules}.
\newblock \emph{\bibinfo{journal}{Nature}} \textbf{\bibinfo{volume}{581}},
  \bibinfo{pages}{396--400} (\bibinfo{year}{2020}).
\newblock \urlprefix\url{https://doi.org/10.1038/s41586-020-2299-4}.

\bibitem{udrescu:2021}
\bibinfo{author}{Udrescu, S.~M.} \emph{et~al.}
\newblock \bibinfo{title}{Isotope shifts of radium monofluoride molecules}.
\newblock \emph{\bibinfo{journal}{Phys. Rev. Lett.}}
  \textbf{\bibinfo{volume}{127}}, \bibinfo{pages}{033001}
  (\bibinfo{year}{2021}).
\newblock \eprint{2105.10549}.

\bibitem{ahmad:1982}
\bibinfo{author}{Ahmad, I.}, \bibinfo{author}{Gindler, J.~E.},
  \bibinfo{author}{Betts, R.~R.}, \bibinfo{author}{Chasman, R.~R.} \&
  \bibinfo{author}{Friedman, A.~M.}
\newblock \bibinfo{title}{Possible ground-state octupole deformation in
  {$^{229}\mathrm{Pa}$}}.
\newblock \emph{\bibinfo{journal}{Phys. Rev. Lett.}}
  \textbf{\bibinfo{volume}{49}}, \bibinfo{pages}{1758--1761}
  (\bibinfo{year}{1982}).
\newblock \urlprefix\url{https://link.aps.org/doi/10.1103/PhysRevLett.49.1758}.

\bibitem{ahmad:2015}
\bibinfo{author}{Ahmad, I.}, \bibinfo{author}{Chasman, R.~R.},
  \bibinfo{author}{Greene, J.~P.}, \bibinfo{author}{Kondev, F.~G.} \&
  \bibinfo{author}{Zhu, S.}
\newblock \bibinfo{title}{Electron capture decay of 58-min
  {$_{92}^{229}\mathrm{U}$} and levels in {$_{91}^{229}\mathrm{Pa}$}}.
\newblock \emph{\bibinfo{journal}{Phys. Rev. C}} \textbf{\bibinfo{volume}{92}},
  \bibinfo{pages}{024313} (\bibinfo{year}{2015}).
\newblock \urlprefix\url{https://link.aps.org/doi/10.1103/PhysRevC.92.024313}.

\bibitem{singh:2019}
\bibinfo{author}{Singh, J.~T.}
\newblock \bibinfo{title}{A new concept for searching for time-reversal
  symmetry violation using {Pa-229} ions trapped in optical crystals}.
\newblock \emph{\bibinfo{journal}{Hyperfine Interact.}}
  \textbf{\bibinfo{volume}{240}}, \bibinfo{pages}{29} (\bibinfo{year}{2019}).
\newblock \urlprefix\url{https://doi.org/10.1007/s10751-019-1573-z}.

\bibitem{auerbach:1996}
\bibinfo{author}{Auerbach, N.}, \bibinfo{author}{Flambaum, V.~V.} \&
  \bibinfo{author}{Spevak, V.}
\newblock \bibinfo{title}{Collective {T-} and {P}-odd electromagnetic moments
  in nuclei with octupole deformations}.
\newblock \emph{\bibinfo{journal}{Phys. Rev. Lett.}}
  \textbf{\bibinfo{volume}{76}}, \bibinfo{pages}{4316--4319}
  (\bibinfo{year}{1996}).
\newblock \urlprefix\url{https://link.aps.org/doi/10.1103/PhysRevLett.76.4316}.

\bibitem{flambaum:2019}
\bibinfo{author}{Flambaum, V.~V.}
\newblock \bibinfo{title}{Enhanced nuclear {Schiff} moment and time-reversal
  violation in {$^{229}\mathrm{Th}$}-containing molecules}.
\newblock \emph{\bibinfo{journal}{Phys. Rev. C}} \textbf{\bibinfo{volume}{99}},
  \bibinfo{pages}{035501} (\bibinfo{year}{2019}).
\newblock \urlprefix\url{https://link.aps.org/doi/10.1103/PhysRevC.99.035501}.

\bibitem{gaul:2020}
\bibinfo{author}{Gaul, K.} \& \bibinfo{author}{Berger, R.}
\newblock \bibinfo{title}{Toolbox approach for quasi-relativistic calculation
  of molecular properties for precision tests of fundamental physics}.
\newblock \emph{\bibinfo{journal}{J. Chem. Phys.}}
  \textbf{\bibinfo{volume}{152}}, \bibinfo{pages}{044101}
  (\bibinfo{year}{2020}).
\newblock \urlprefix\url{https://doi.org/10.1063/1.5121483}.
\newblock \eprint{1907.10432}.

\bibitem{hinds:1980}
\bibinfo{author}{Hinds, E.~A.} \& \bibinfo{author}{Sandars, P. G.~H.}
\newblock \bibinfo{title}{Electric dipole hyperfine structure of {TlF}}.
\newblock \emph{\bibinfo{journal}{Phys. Rev. A}} \textbf{\bibinfo{volume}{21}},
  \bibinfo{pages}{471--479} (\bibinfo{year}{1980}).
\newblock \urlprefix\url{https://link.aps.org/doi/10.1103/PhysRevA.21.471}.

\bibitem{kozlov:1995}
\bibinfo{author}{Kozlov, M.~G.} \& \bibinfo{author}{Labzowsky, L.~N.}
\newblock \bibinfo{title}{Parity violation effects in diatomics}.
\newblock \emph{\bibinfo{journal}{J. Phys. B}} \textbf{\bibinfo{volume}{28}},
  \bibinfo{pages}{1933--1961} (\bibinfo{year}{1995}).

\bibitem{aggarwal:2018}
\bibinfo{author}{Aggarwal, P.} \emph{et~al.}
\newblock \bibinfo{title}{Measuring the electric dipole moment of the electron
  in {BaF}}.
\newblock \emph{\bibinfo{journal}{Eur. Phys. J. D}}
  \textbf{\bibinfo{volume}{72}}, \bibinfo{pages}{197} (\bibinfo{year}{2018}).
\newblock \urlprefix\url{https://doi.org/10.1140/epjd/e2018-90192-9}.

\bibitem{gaul:2019}
\bibinfo{author}{Gaul, K.}, \bibinfo{author}{Marquardt, S.},
  \bibinfo{author}{Isaev, T.} \& \bibinfo{author}{Berger, R.}
\newblock \bibinfo{title}{{Systematic study of relativistic and chemical
  enhancements of $\mathcal{P},\mathcal{T}$-odd effects in polar diatomic
  radicals}}.
\newblock \emph{\bibinfo{journal}{Phys. Rev. A}} \textbf{\bibinfo{volume}{99}},
  \bibinfo{pages}{032509} (\bibinfo{year}{2019}).
\newblock \urlprefix\url{https://link.aps.org/doi/10.1103/PhysRevA.99.032509}.
\newblock \eprint{1805.05494}.

\bibitem{chin:2009}
\bibinfo{author}{Chin, C.}, \bibinfo{author}{Flambaum, V.~V.} \&
  \bibinfo{author}{Kozlov, M.~G.}
\newblock \bibinfo{title}{Ultracold molecules: new probes on the variation of
  fundamental constants}.
\newblock \emph{\bibinfo{journal}{New J. Phys.}} \textbf{\bibinfo{volume}{11}},
  \bibinfo{pages}{055048} (\bibinfo{year}{2009}).
\newblock \urlprefix\url{https://doi.org/10.1088/1367-2630/11/5/055048}.

\bibitem{kovacs:2013}
\bibinfo{author}{Kov{\'a}cs, A.}, \bibinfo{author}{Infante, I.} \&
  \bibinfo{author}{Gagliardi, L.}
\newblock \bibinfo{title}{Theoretic study of the electronic spectra of neutral
  and cationic {PaO} and {PaO$_2$}}.
\newblock \emph{\bibinfo{journal}{Struct. Chem.}}
  \textbf{\bibinfo{volume}{24}}, \bibinfo{pages}{917--925}
  (\bibinfo{year}{2013}).
\newblock \urlprefix\url{https://doi.org/10.1007/s11224-013-0251-z}.

\bibitem{santos:2006}
\bibinfo{author}{Santos, M.} \emph{et~al.}
\newblock \bibinfo{title}{Oxidation of gas-phase protactinium ions, {Pa$^+$ and
  Pa$^{2+}$}: Formation and properties of {PaO$_2^{2+}$(g), P}rotactinyl}.
\newblock \emph{\bibinfo{journal}{J. Phys. Chem. A}}
  \textbf{\bibinfo{volume}{110}}, \bibinfo{pages}{5751--5759}
  (\bibinfo{year}{2006}).
\newblock \urlprefix\url{https://doi.org/10.1021/jp057297d}.
\newblock \bibinfo{note}{PMID: 16640369},
  \eprint{https://doi.org/10.1021/jp057297d}.

\bibitem{isaev:2016}
\bibinfo{author}{Isaev, T.~A.} \& \bibinfo{author}{Berger, R.}
\newblock \bibinfo{title}{Polyatomic candidates for cooling of molecules with
  lasers from simple theoretical concepts}.
\newblock \emph{\bibinfo{journal}{Phys. Rev. Lett.}}
  \textbf{\bibinfo{volume}{116}}, \bibinfo{pages}{063006}
  (\bibinfo{year}{2016}).
\newblock
  \urlprefix\url{https://link.aps.org/doi/10.1103/PhysRevLett.116.063006}.

\bibitem{kozyryev:2016}
\bibinfo{author}{Kozyryev, I.}, \bibinfo{author}{Baum, L.},
  \bibinfo{author}{Matsuda, K.} \& \bibinfo{author}{Doyle, J.~M.}
\newblock \bibinfo{title}{Proposal for laser cooling of complex polyatomic
  molecules}.
\newblock \emph{\bibinfo{journal}{ChemPhysChem}} \textbf{\bibinfo{volume}{17}},
  \bibinfo{pages}{3641--3648} (\bibinfo{year}{2016}).
\newblock
  \urlprefix\url{https://onlinelibrary.wiley.com/doi/abs/10.1002/cphc.201601051}.

\bibitem{fan:2021}
\bibinfo{author}{Fan, M.} \emph{et~al.}
\newblock \bibinfo{title}{Optical mass spectrometry of cold
  {${\mathrm{RaOH}}^{+}$} and {${{\mathrm{RaOCH}}_{3}}^{+}$}}.
\newblock \emph{\bibinfo{journal}{Phys. Rev. Lett.}}
  \textbf{\bibinfo{volume}{126}}, \bibinfo{pages}{023002}
  (\bibinfo{year}{2021}).
\newblock
  \urlprefix\url{https://link.aps.org/doi/10.1103/PhysRevLett.126.023002}.

\bibitem{isaev:2017}
\bibinfo{author}{Isaev, T.~A.}, \bibinfo{author}{Zaitsevskii, A.~V.} \&
  \bibinfo{author}{Eliav, E.}
\newblock \bibinfo{title}{Laser-coolable polyatomic molecules with heavy
  nuclei}.
\newblock \emph{\bibinfo{journal}{J. Phys. B}} \textbf{\bibinfo{volume}{50}},
  \bibinfo{pages}{225101} (\bibinfo{year}{2017}).
\newblock \urlprefix\url{http://stacks.iop.org/0953-4075/50/i=22/a=225101}.

\bibitem{kozyryev:2017a}
\bibinfo{author}{Kozyryev, I.} \& \bibinfo{author}{Hutzler, N.~R.}
\newblock \bibinfo{title}{Precision measurement of time-reversal symmetry
  violation with laser-cooled polyatomic molecules}.
\newblock \emph{\bibinfo{journal}{Phys. Rev. Lett.}}
  \textbf{\bibinfo{volume}{119}}, \bibinfo{pages}{133002}
  (\bibinfo{year}{2017}).
\newblock
  \urlprefix\url{https://link.aps.org/doi/10.1103/PhysRevLett.119.133002}.

\bibitem{harvey:2003}
\bibinfo{author}{Harvey, J.~N.} \& \bibinfo{author}{Kaczorowska, M.}
\newblock \bibinfo{title}{Microsolvation of metal ions: on the stability of
  {[Zr(CH$_3$CN)]$^{4+}$} and other multiply charged ions}.
\newblock \emph{\bibinfo{journal}{Int. J. Mass Spec.}}
  \textbf{\bibinfo{volume}{228}}, \bibinfo{pages}{517--526}
  (\bibinfo{year}{2003}).
\newblock
  \urlprefix\url{https://www.sciencedirect.com/science/article/pii/S138738060300160X}.
\newblock \bibinfo{note}{Special Issue: In honour of Helmut Schwarz}.

\bibitem{dirac19}
\bibinfo{note}{{DIRAC}, a relativistic ab initio electronic structure program,
  Release {DIRAC19} (2019), written by A.~S.~P.~Gomes, T.~Saue, L.~Visscher,
  H.~J.~{\relax Aa}.~Jensen, and R.~Bast, with contributions from I.~A.~Aucar,
  V.~Bakken, K.~G.~Dyall, S.~Dubillard, U.~Ekstr{\"o}m, E.~Eliav,
  T.~Enevoldsen, E.~Fa{\ss}hauer, T.~Fleig, O.~Fossgaard, L.~Halbert,
  E.~D.~Hedeg{\aa}rd, B.~Heimlich--Paris, T.~Helgaker, J.~Henriksson,
  M.~Ilia{\v{s}}, Ch.~R.~Jacob, S.~Knecht, S.~Komorovsk{\'y}, O.~Kullie,
  J.~K.~L{\ae}rdahl, C.~V.~Larsen, Y.~S.~Lee, H.~S.~Nataraj, M.~K.~Nayak,
  P.~Norman, G.~Olejniczak, J.~Olsen, J.~M.~H.~Olsen, Y.~C.~Park,
  J.~K.~Pedersen, M.~Pernpointner, R.~di~Remigio, K.~Ruud, P.~Sa{\l}ek,
  B.~Schimmelpfennig, B.~Senjean, A.~Shee, J.~Sikkema, A.~J.~Thorvaldsen,
  J.~Thyssen, J.~van~Stralen, M.~L.~Vidal, S.~Villaume, O.~Visser, T.~Winther,
  and S.~Yamamoto (available at \url{http://dx.doi.org/10.5281/zenodo.3572669},
  see also \url{http://www.diracprogram.org})}.

\bibitem{dyall:2002}
\bibinfo{author}{Dyall, K.~G.}
\newblock \bibinfo{title}{Relativistic and nonrelativistic finite nucleus
  optimized triple-zeta basis sets for the 4p, 5p and 6p elements}.
\newblock \emph{\bibinfo{journal}{Theor. Chem. Acc.}}
  \textbf{\bibinfo{volume}{108}}, \bibinfo{pages}{335--340}
  (\bibinfo{year}{2002}).
\newblock \urlprefix\url{http://dx.doi.org/10.1007/s00214-002-0388-0}.

\bibitem{dyall:2006}
\bibinfo{author}{Dyall, K.~G.}
\newblock \bibinfo{title}{Relativistic quadruple-zeta and revised triple-zeta
  and double-zeta basis sets for the 4p, 5p, and 6p elements}.
\newblock \emph{\bibinfo{journal}{Theor. Chem. Acc.}}
  \textbf{\bibinfo{volume}{115}}, \bibinfo{pages}{441--447}
  (\bibinfo{year}{2006}).
\newblock \urlprefix\url{http://dx.doi.org/10.1007/s00214-006-0126-0}.

\bibitem{roos:2005}
\bibinfo{author}{Roos, B.~O.}, \bibinfo{author}{Lindh, R.},
  \bibinfo{author}{Malmqvist, P.}, \bibinfo{author}{Veryazov, V.} \&
  \bibinfo{author}{Widmark, P.-O.}
\newblock \bibinfo{title}{New relativistic {ANO} basis sets for actinide
  atoms}.
\newblock \emph{\bibinfo{journal}{Chem. Phys. Lett.}}
  \textbf{\bibinfo{volume}{409}}, \bibinfo{pages}{295--299}
  (\bibinfo{year}{2005}).
\newblock
  \urlprefix\url{https://www.sciencedirect.com/science/article/pii/S0009261405006810}.

\bibitem{wullen:2010}
\bibinfo{author}{van W{\"{u}}llen, C.}
\newblock \bibinfo{title}{{A Quasirelativistic Two-component Density Functional
  and Hartree-Fock Program}}.
\newblock \emph{\bibinfo{journal}{Z. Phys. Chem}}
  \textbf{\bibinfo{volume}{224}}, \bibinfo{pages}{413--426}
  (\bibinfo{year}{2010}).

\bibitem{ahlrichs:1989}
\bibinfo{author}{Ahlrichs, R.}, \bibinfo{author}{B{\"a}r, M.},
  \bibinfo{author}{H{\"a}ser, M.}, \bibinfo{author}{Horn, H.} \&
  \bibinfo{author}{K{\"o}lmel, C.}
\newblock \bibinfo{title}{Electronic structure calculations on workstation
  computers: {T}he program system turbomole}.
\newblock \emph{\bibinfo{journal}{Chem. Phys. Lett.}}
  \textbf{\bibinfo{volume}{162}}, \bibinfo{pages}{165--169}
  (\bibinfo{year}{1989}).

\bibitem{wullen:1998}
\bibinfo{author}{van W{\"u}llen, C.}
\newblock \bibinfo{title}{Molecular density functional calculations in the
  regular relativistic approximation: {M}ethod, application to coinage metal
  diatomics, hydrides, fluorides and chlorides, and comparison with first-order
  relativistic calculations}.
\newblock \emph{\bibinfo{journal}{J. Chem. Phys.}}
  \textbf{\bibinfo{volume}{109}}, \bibinfo{pages}{392--399}
  (\bibinfo{year}{1998}).

\bibitem{liu:2002}
\bibinfo{author}{Liu, W.}, \bibinfo{author}{van W{\"u}llen, C.},
  \bibinfo{author}{Wang, F.} \& \bibinfo{author}{Li, L.}
\newblock \bibinfo{title}{Spectroscopic constants of {MH} and {M}$_2$ ({M} =
  {Tl}, {E113}, {Bi}, {E115}): direct comparisons of four- and two-component
  approaches in the framework of relativistic density functional theory}.
\newblock \emph{\bibinfo{journal}{J. Chem. Phys.}}
  \textbf{\bibinfo{volume}{116}}, \bibinfo{pages}{3626--3634}
  (\bibinfo{year}{2002}).

\bibitem{vosko:1980}
\bibinfo{author}{Vosko, S.~H.}, \bibinfo{author}{Wilk, L.} \&
  \bibinfo{author}{Nuisar, M.}
\newblock \bibinfo{title}{Accurate spin-dependent electron liquid correlation
  energies for local spin density calculations: {A} critical analysis}.
\newblock \emph{\bibinfo{journal}{Can. J. Phys.}}
  \textbf{\bibinfo{volume}{58}}, \bibinfo{pages}{1200--1211}
  (\bibinfo{year}{1980}).

\bibitem{becke:1988}
\bibinfo{author}{Becke, A.~D.}
\newblock \bibinfo{title}{Density-functional exchange-energy approximation with
  correct asymptotic-behavior}.
\newblock \emph{\bibinfo{journal}{Phys. Rev. A}} \textbf{\bibinfo{volume}{38}},
  \bibinfo{pages}{3098--3100} (\bibinfo{year}{1988}).

\bibitem{lee:1988}
\bibinfo{author}{Lee, C.}, \bibinfo{author}{Yang, W.} \& \bibinfo{author}{Parr,
  R.~G.}
\newblock \bibinfo{title}{Development of the {C}olle-{S}alvetti
  correlation-energy formula into a functional of the electron-density}.
\newblock \emph{\bibinfo{journal}{Phys. Rev. B}} \textbf{\bibinfo{volume}{37}},
  \bibinfo{pages}{785--789} (\bibinfo{year}{1988}).

\bibitem{stephens:1994}
\bibinfo{author}{Stephens, P.~J.}, \bibinfo{author}{Devlin, F.~J.},
  \bibinfo{author}{Chabalowski, C.~F.} \& \bibinfo{author}{Frisch, M.~J.}
\newblock \bibinfo{title}{Ab initio calculation of vibrational absorption and
  circular dichroism spectra using density functional force fields}.
\newblock \emph{\bibinfo{journal}{J. Phys. Chem.}}
  \textbf{\bibinfo{volume}{98}}, \bibinfo{pages}{11623--11627}
  (\bibinfo{year}{1994}).

\bibitem{perdew:1996}
\bibinfo{author}{Perdew, J.~P.}, \bibinfo{author}{Burke, K.} \&
  \bibinfo{author}{Ernzerhof, M.}
\newblock \bibinfo{title}{Generalized gradient approximation made simple}.
\newblock \emph{\bibinfo{journal}{Phys. Rev. Lett.}}
  \textbf{\bibinfo{volume}{77}}, \bibinfo{pages}{3865--3868}
  (\bibinfo{year}{1996}).

\bibitem{adamo:1999}
\bibinfo{author}{Adamo, C.} \& \bibinfo{author}{Barone, V.}
\newblock \bibinfo{title}{Toward reliable density functional methods without
  adjustable parameters: The {PBE0} model}.
\newblock \emph{\bibinfo{journal}{J. Chem. Phys.}}
  \textbf{\bibinfo{volume}{110}}, \bibinfo{pages}{6158--6170}
  (\bibinfo{year}{1999}).
\newblock \urlprefix\url{https://doi.org/10.1063/1.478522}.

\bibitem{roos:2004}
\bibinfo{author}{Roos, B.~O.}, \bibinfo{author}{Lindh, R.},
  \bibinfo{author}{Malmqvist, P.}, \bibinfo{author}{Veryazov, V.} \&
  \bibinfo{author}{Widmark, P.~O.}
\newblock \bibinfo{title}{{Main Group Atoms and Dimers Studied with a New
  Relativistic ANO Basis Set}}.
\newblock \emph{\bibinfo{journal}{J. Phys. Chem. A}}
  \textbf{\bibinfo{volume}{108}}, \bibinfo{pages}{2851--2858}
  (\bibinfo{year}{2004}).

\bibitem{gaul:2017}
\bibinfo{author}{Gaul, K.} \& \bibinfo{author}{Berger, R.}
\newblock \bibinfo{title}{Zeroth order regular approximation approach to
  electric dipole moment interactions of the electron}.
\newblock \emph{\bibinfo{journal}{J. Chem. Phys.}}
  \textbf{\bibinfo{volume}{147}}, \bibinfo{pages}{014109}
  (\bibinfo{year}{2017}).

\bibitem{visscher:1997}
\bibinfo{author}{Visscher, L.} \& \bibinfo{author}{Dyall, K.~G.}
\newblock \bibinfo{title}{Dirac-fock atomic electronic structure calculations
  using different nuclear charge distributions}.
\newblock \emph{\bibinfo{journal}{At. Data Nucl. Data Tables}}
  \textbf{\bibinfo{volume}{67}}, \bibinfo{pages}{207--224}
  (\bibinfo{year}{1997}).

\bibitem{gilbert:2008}
\bibinfo{author}{Gilbert, A. T.~B.}, \bibinfo{author}{Besley, N.~A.} \&
  \bibinfo{author}{Gill, P. M.~W.}
\newblock \bibinfo{title}{Self-consistent field calculations of excited states
  using the maximum overlap method {(MOM)}}.
\newblock \emph{\bibinfo{journal}{J. Phys. Chem. A}}
  \textbf{\bibinfo{volume}{112}}, \bibinfo{pages}{13164--13171}
  (\bibinfo{year}{2008}).
\newblock \urlprefix\url{https://doi.org/10.1021/jp801738f}.
\newblock \bibinfo{note}{PMID: 18729344},
  \eprint{https://doi.org/10.1021/jp801738f}.

\bibitem{barca:2018}
\bibinfo{author}{Barca, G. M.~J.}, \bibinfo{author}{Gilbert, A. T.~B.} \&
  \bibinfo{author}{Gill, P. M.~W.}
\newblock \bibinfo{title}{Simple models for difficult electronic excitations}.
\newblock \emph{\bibinfo{journal}{J. Chem. Theo. Comp.}}
  \textbf{\bibinfo{volume}{14}}, \bibinfo{pages}{1501--1509}
  (\bibinfo{year}{2018}).
\newblock \urlprefix\url{https://doi.org/10.1021/acs.jctc.7b00994}.
\newblock \bibinfo{note}{PMID: 29444408},
  \eprint{https://doi.org/10.1021/acs.jctc.7b00994}.

\bibitem{berger:1997a}
\bibinfo{author}{Berger, R.}, \bibinfo{author}{Fischer, C.} \&
  \bibinfo{author}{Klessinger, M.}
\newblock \bibinfo{title}{Calculation of the vibronic fine structure in
  electronic spectra at higher temperatures. 1. benzene and pyrazine}.
\newblock \emph{\bibinfo{journal}{J. Phys. Chem. A}}
  \textbf{\bibinfo{volume}{102}}, \bibinfo{pages}{7157--7167}
  (\bibinfo{year}{1998}).

\bibitem{jankowiak:2007}
\bibinfo{author}{Jankowiak, H.-C.}, \bibinfo{author}{Stuber, J.~L.} \&
  \bibinfo{author}{Berger, R.}
\newblock \bibinfo{title}{Vibronic transitions in large molecular systems:
  Rigorous prescreening conditions for {F}ranck-{C}ondon factors}.
\newblock \emph{\bibinfo{journal}{J. Chem. Phys.}}
  \textbf{\bibinfo{volume}{127}}, \bibinfo{pages}{234101}
  (\bibinfo{year}{2007}).

\bibitem{huh:2012proc}
\bibinfo{author}{Huh, J.} \& \bibinfo{author}{Berger, R.}
\newblock \bibinfo{title}{Coherent state-based generating function approach for
  {F}ranck-{C}ondon transitions and beyond}.
\newblock In \emph{\bibinfo{booktitle}{{SYMMETRIES IN SCIENCE XV}}}, vol.
  \bibinfo{volume}{380} of \emph{\bibinfo{series}{J. Phys. Conf. Ser.}}
  (\bibinfo{year}{2012}).
\newblock \bibinfo{note}{International Symposium on Symmetries in Science XV,
  Bregenz, AUSTRIA, JUL 31-AUG 05, 2011}.

\bibitem{huh:2010}
\bibinfo{author}{Huh, J.}, \bibinfo{author}{Neff, M.}, \bibinfo{author}{Rauhut,
  G.} \& \bibinfo{author}{Berger, R.}
\newblock \bibinfo{title}{{F}ranck-{C}ondon profiles in
  photodetachment-photoelectron spectra of $\mathrm{HS}^{-}_{2}$ and
  $\mathrm{DS}^{-}_{2}$ based on vibrational configuration interaction
  wavefunctions}.
\newblock \emph{\bibinfo{journal}{Mol. Phys.}} \textbf{\bibinfo{volume}{108}},
  \bibinfo{pages}{409} (\bibinfo{year}{2010}).

\bibitem{Lowdin:55}
\bibinfo{author}{L{\"o}wdin, P.-O.}
\newblock \bibinfo{title}{Quantum theory of many-particle systems .1. physical
  interpretations by means of density matrices, natural spin-orbitals, and
  convergence problems in the method of configurational interaction}.
\newblock \emph{\bibinfo{journal}{Phys. Rev.}} \textbf{\bibinfo{volume}{97}},
  \bibinfo{pages}{1474--1489} (\bibinfo{year}{1955}).

\bibitem{flambaum:2002}
\bibinfo{author}{Flambaum, V.~V.} \& \bibinfo{author}{Ginges, J. S.~M.}
\newblock \bibinfo{title}{Nuclear {Schiff} moment and time-invariance violation
  in atoms}.
\newblock \emph{\bibinfo{journal}{Phys. Rev. A}} \textbf{\bibinfo{volume}{65}},
  \bibinfo{pages}{032113} (\bibinfo{year}{2002}).
\newblock \urlprefix\url{https://link.aps.org/doi/10.1103/PhysRevA.65.032113}.

\bibitem{flambaum:2020a}
\bibinfo{author}{Flambaum, V.~V.}, \bibinfo{author}{Dzuba, V.~A.} \&
  \bibinfo{author}{Tran~Tan, H.~B.}
\newblock \bibinfo{title}{Time- and parity-violating effects of the nuclear
  schiff moment in molecules and solids}.
\newblock \emph{\bibinfo{journal}{Phys. Rev. A}}
  \textbf{\bibinfo{volume}{101}}, \bibinfo{pages}{042501}
  (\bibinfo{year}{2020}).
\newblock \urlprefix\url{https://link.aps.org/doi/10.1103/PhysRevA.101.042501}.

\bibitem{molpro2012a}
\bibinfo{author}{Werner, H.-J.}, \bibinfo{author}{Knowles, P.~J.},
  \bibinfo{author}{Knizia, G.}, \bibinfo{author}{Manby, F.~R.} \&
  \bibinfo{author}{Schütz, M.}
\newblock \bibinfo{title}{Molpro: a general-purpose quantum chemistry program
  package}.
\newblock \emph{\bibinfo{journal}{WIREs Computational Molecular Science}}
  \textbf{\bibinfo{volume}{2}}, \bibinfo{pages}{242--253}
  (\bibinfo{year}{2012}).
\newblock
  \urlprefix\url{https://wires.onlinelibrary.wiley.com/doi/abs/10.1002/wcms.82}.

\bibitem{molpro2019a}
\bibinfo{author}{Werner, H.-J.} \emph{et~al.}
\newblock \bibinfo{title}{Molpro, version 2019.2, a package of ab initio
  programs} (\bibinfo{year}{2019}).
\newblock \bibinfo{note}{See http://www.molpro.net}.

\bibitem{molpro2020}
\bibinfo{author}{Werner, H.-J.} \emph{et~al.}
\newblock \bibinfo{title}{The molpro quantum chemistry package}.
\newblock \emph{\bibinfo{journal}{J. Chem. Phys.}}
  \textbf{\bibinfo{volume}{152}}, \bibinfo{pages}{144107}
  (\bibinfo{year}{2020}).
\newblock \urlprefix\url{https://doi.org/10.1063/5.0005081}.

\bibitem{cao:2003b}
\bibinfo{author}{Cao, X.}, \bibinfo{author}{Dolg, M.} \&
  \bibinfo{author}{Stoll, H.}
\newblock \bibinfo{title}{Valence basis sets for relativistic energy-consistent
  small-core actinide pseudopotentials}.
\newblock \emph{\bibinfo{journal}{J. Chem. Phys.}}
  \textbf{\bibinfo{volume}{118}}, \bibinfo{pages}{487--496}
  (\bibinfo{year}{2003}).
\newblock \urlprefix\url{https://doi.org/10.1063/1.1521431}.

\bibitem{dunning:1989}
\bibinfo{author}{Dunning, T.~H., Jr.}
\newblock \bibinfo{title}{Gaussian basis sets for use in correlated molecular
  calculations. i. the atoms boron through neon and hydrogen}.
\newblock \emph{\bibinfo{journal}{J. Chem. Phys.}}
  \textbf{\bibinfo{volume}{90}}, \bibinfo{pages}{1007--1023}
  (\bibinfo{year}{1989}).

\bibitem{mathematica11}
\bibinfo{author}{{Wolfram Research, Inc.}}
\newblock \bibinfo{title}{{Mathematica 11.0}} (\bibinfo{year}{2016}).

\bibitem{batista:2004}
\bibinfo{author}{Batista, E.~R.}, \bibinfo{author}{Martin, R.~L.} \&
  \bibinfo{author}{Hay, P.~J.}
\newblock \bibinfo{title}{Density functional investigations of the properties
  and thermochemistry of {UF$_n$} and {UCl$_n$} {$(n=1,\dots,6)$}}.
\newblock \emph{\bibinfo{journal}{J. Chem. Phys.}}
  \textbf{\bibinfo{volume}{121}}, \bibinfo{pages}{11104--11111}
  (\bibinfo{year}{2004}).
\newblock \urlprefix\url{https://aip.scitation.org/doi/abs/10.1063/1.1811607}.
\newblock \eprint{https://aip.scitation.org/doi/pdf/10.1063/1.1811607}.

\bibitem{blondel:2001}
\bibinfo{author}{Blondel, C.}, \bibinfo{author}{Delsart, C.} \&
  \bibinfo{author}{Goldfarb, F.}
\newblock \bibinfo{title}{Electron spectrometry at the $\mu$ev level and the
  electron affinities of si and f}.
\newblock \emph{\bibinfo{journal}{J. Phys. B: Atom. Mol. Opt. Phys.}}
  \textbf{\bibinfo{volume}{34}}, \bibinfo{pages}{L281--L288}
  (\bibinfo{year}{2001}).
\newblock \urlprefix\url{http://stacks.iop.org/0953-4075/34/i=9/a=101}.

\bibitem{walker:2001}
\bibinfo{author}{Walker, N.~R.}, \bibinfo{author}{Wright, R.~R.},
  \bibinfo{author}{Barran, P.~E.}, \bibinfo{author}{Murrell, J.~N.} \&
  \bibinfo{author}{Stace, A.~J.}
\newblock \bibinfo{title}{Comparisons in the behavior of stable {Copper(II),
  Silver(II), and Gold(II)} complexes in the gas phase: Are there
  implications for condensed-phase chemistry?}
\newblock \emph{\bibinfo{journal}{J. Am. Chem. Soc.}}
  \textbf{\bibinfo{volume}{123}}, \bibinfo{pages}{4223--4227}
  (\bibinfo{year}{2001}).
\newblock \urlprefix\url{https://doi.org/10.1021/ja003431q}.
\newblock \bibinfo{note}{PMID: 11457187}.

\bibitem{axe:1961}
\bibinfo{author}{Axe, J.~D.}, \bibinfo{author}{Stapleton, H.~J.} \&
  \bibinfo{author}{Jeffries, C.~D.}
\newblock \bibinfo{title}{Paramagnetic resonance hyperfine structure of
  tetravalent {${\mathrm{Pa}}^{231}$ in
  ${\mathrm{Cs}}_{2}$Zr${\mathrm{Cl}}_{6}$}}.
\newblock \emph{\bibinfo{journal}{Phys. Rev.}} \textbf{\bibinfo{volume}{121}},
  \bibinfo{pages}{1630--1637} (\bibinfo{year}{1961}).
\newblock \urlprefix\url{https://link.aps.org/doi/10.1103/PhysRev.121.1630}.

\end{thebibliography}
\end{document}